\documentclass[11pt]{elsarticle}
\usepackage[margin=1in]{geometry}
\usepackage{natbib}
\usepackage{adjustbox}
\usepackage{graphicx}
\usepackage{mathrsfs}
\usepackage{color}
\usepackage{amssymb}
\usepackage{amsfonts}
\usepackage{amsmath}
\usepackage{bbm}
\usepackage{nameref}
\usepackage{tikz}
\usepackage{stackengine}
\usetikzlibrary{bayesnet}
\usepackage{lineno}
\usetikzlibrary{decorations.pathreplacing}
\usepackage{setspace}

\DeclareFontFamily{U}{rcjhbltx}{}
\DeclareFontShape{U}{rcjhbltx}{m}{n}{<->rcjhbltx}{}
\DeclareSymbolFont{hebrewletters}{U}{rcjhbltx}{m}{n}

\let\aleph\relax\let\beth\relax
\let\gimel\relax\let\daleth\relax

\DeclareMathSymbol{\aleph}{\mathord}{hebrewletters}{39}
\DeclareMathSymbol{\beth}{\mathord}{hebrewletters}{98}
\DeclareMathSymbol{\gimel}{\mathord}{hebrewletters}{103}
\DeclareMathSymbol{\daleth}{\mathord}{hebrewletters}{100}
\DeclareMathSymbol{\lamed}{\mathord}{hebrewletters}{108}
\DeclareMathSymbol{\mem}{\mathord}{hebrewletters}{109}
\DeclareMathSymbol{\ayin}{\mathord}{hebrewletters}{96}
\DeclareMathSymbol{\tsadi}{\mathord}{hebrewletters}{118}
\DeclareMathSymbol{\qof}{\mathord}{hebrewletters}{113}
\DeclareMathSymbol{\shin}{\mathord}{hebrewletters}{152}

\usepackage{hyperref}
\allowdisplaybreaks

\usepackage{subcaption}

\usepackage{longtable}

\usepackage{algorithm}
\usepackage{algpseudocode}

\usepackage{amsthm}

\theoremstyle{definition}

\newtheorem*{definition*}{Definition}

\algtext*{EndFor}% Remove "end for" text
\algtext*{EndIf}% Remove "end if" text
\algtext*{EndFunction}% Remove "end function" text
\algtext*{EndWhile}% Remove "end while" text

\algrenewcommand\algorithmicforall{\textbf{foreach}}
\algrenewcommand\algorithmicindent{.8em}

\usepackage{placeins}

\begin{document}
%\journal{Transportation Research Part C: Emerging Technologies}
\begin{frontmatter}

\title{Design of Transit-Centric Multimodal Urban Mobility System with Autonomous Mobility-on-Demand}

\author[mitcee]{Xiaotong Guo}
\ead{xtguo@mit.edu}
\author[dusp]{Jinhua Zhao\corref{cor}}
\ead{jinhua@mit.edu}

\address[mitcee]{Department of Civil and Environmental Engineering, Massachusetts Institute of Technology, Cambridge, MA 02139, USA}
\address[dusp]{Department of Urban Studies and Planning, Massachusetts Institute of Technology, Cambridge, MA 02139, USA}
\cortext[cor]{Corresponding author}

\begin{abstract}

This paper addresses the pressing challenge of urban mobility in the context of growing urban populations, changing demand patterns for urban mobility, and emerging technologies like Mobility-on-Demand (MoD) platforms and Autonomous Vehicle (AV). As urban areas swell and demand pattern changes, the integration of Autonomous Mobility-on-Demand (AMoD) systems with existing public transit (PT) networks presents great opportunities to enhancing urban mobility. We propose a novel optimization framework for solving the Transit-Centric Multimodal Urban Mobility with Autonomous Mobility-on-Demand (TCMUM-AMoD) at scale. The system operator (public transit agency) determines the network design and frequency settings of the PT network, fleet sizing and allocations of AMoD system, and the pricing for using the multimodal system with the goal of minimizing passenger disutility. Passengers' mode and route choice behaviors are modelled explicitly using discrete choice models. A first-order approximation algorithm is introduced to solve the problem at scale. Using a case study in Chicago, we showcase the potential to optimize urban mobility across different demand scenarios. To our knowledge, ours is the first paper to jointly optimize transit network design, fleet sizing, and pricing for the multimodal mobility system while considering passengers' mode and route choices.

\end{abstract}

\begin{keyword}
Multimodal Mobility System, Autonomous Mobility-on-Demand, Transit Network Design, Fleet Sizing, Pricing, Nonlinear Optimization, First-Order Approximation
\end{keyword}

\end{frontmatter}

% \linenumbers

\section{Introduction}

The global population is increasingly urban-centric, with 51\%, or 3.5 billion people, currently residing in cities. The proportion is projected to rise to 70\%, or 6.3 billion people, by the year 2050.  It's anticipated that by 2050, the distances traveled within urban areas will triple~\citep{future_urban_mobility}. Meanwhile, urban mobility demand patterns are undergoing changes in the post-pandemic world, influenced by the increased prevalence of remote working. One of the most significant challenges urban areas are confronting is urban mobility. With urban areas witnessing both a rise and transformation in travel demand, the necessity for an analytical framework capable of generating a high-capacity urban mobility system becomes crucial. Such a system should not only be efficient but also minimize environmental impact.

Emissions from urban mobility systems are one of the main sources of global warming as transportation currently accounts for 29\% of U.S. Greenhouse Gas Emissions~\citep{transportation_emission}. Building a sustainable urban mobility system is an essential strategy to address climate change-related issues. Emerging Mobility-on-Demand (MoD) services, such as ride-hailing platforms like Uber and Lyft, have introduced responsive and reliable travel options for individuals. However, public transit (PT) systems continue to serve as the foundation of sustainable urban mobility, facilitating efficient city-wide travel for large populations. Both MoD and PT have their unique advantages and challenges. MoD offers flexible and direct services to a few passengers at a higher cost, while PT provides cost-effective transportation for large groups of people. However, the fixed schedules of public transit and concerns about accessibility are significant considerations for passengers.

Autonomous Vehicles (AV) signify a transformative future for urban mobility, offering prospects to enhance the quality of public transportation. Autonomous Mobility-on-Demand (AMoD) systems, such as Waymo and Cruise, have demonstrated significant potential in providing reliable and efficient services to passengers in major U.S. cities. While some researchers contend that AMoD and PT are in competition~\citep{Mo2021}, there is a growing consensus that the two systems could actually complement each other. This paper introduces a novel framework for creating a \emph{Transit-Centric Multimodal Urban Mobility (TCMUM)} system, marking a first step towards the complete integration of AMoD systems and PT networks. Within the TCMUM system we envision, AMoD serves as the cornerstone for first-mile-last-mile (FMLM) services, linking passengers to transit stations. This proposed TCMUM-AMoD system merges the advantages of both transit and Mobility-on-Demand, enhancing accessibility while preserving high capacity and efficient passenger movement.

The TCMUM-AMoD system provides passengers with three transportation options: rail, bus, and AMoD. The design of the TCMUM-AMoD system involves several key components: (i) setting the operational frequency for the rail network, (ii) designing the bus network and determining its frequency, (iii) allocating vehicles for the AMoD fleet, and (iv) determining pricing structures using the TCMUM-AMoD system. Leveraging the existing transit infrastructure, the implementation of the TCMUM-AMoD system could offer numerous enhancements to urban mobility, including: (i) the substitution of infrequent bus services with AMoD fleets, improving accessibility, (ii) the enhancement of bus service frequencies along key transit corridors, reducing passenger wait times and elevating service levels, (iii) improved coordination among different modes of transportation, and (iv) a decrease in long-haul MoD trips, potentially alleviating traffic congestion and lowering carbon emissions.

% \begin{itemize}
%     \item The substitution of infrequent bus services with AMoD fleets, improving accessibility.
%     \item The enhancement of bus service frequencies along key transit corridors, reducing passenger wait times and elevating service levels.
%     \item Improved coordination among different modes of transportation.
%     \item A decrease in long-haul MoD trips, potentially alleviating traffic congestion and lowering carbon emissions.
% \end{itemize}

% Contribution
This paper presents a tractable optimization framework that is designed to simultaneously configure networks and set service prices, with the goal of minimizing the total disutility experienced by passengers within the system. The contribution of this paper can be summarized as follows:

\begin{enumerate}
    \item To the best of authors' knowledge, this work is the first attempt to propose a tractable optimization framework for solving the joint transit network design, fleet sizing, and pricing in the multimodal mobility system while considering passengers' mode and route choices.  
    \item An optimization model for solving the design of Transit-Centric Multimodal Urban Mobility with Autonomous Mobility-on-Demand (TCMUM-AMoD) is proposed, where the sharing scenarios of AMoD services are modelled explicitly and passengers' mode and route choice behaviors are captured by discrete choice models.
    \item The proposed optimization model is challenging to solve due to the non-linearity brought by discrete choice models and the growth of feasible combinations of modes and routes. A first-order approximation algorithm is introduced to solve the problem at scale.
    \item The suggested optimization framework is assessed through a real-world case study in Chicago. It evaluates two types of demand -- local and downtown -- to demonstrate the optimal system design across various demand scenarios.
\end{enumerate}

The remainder of the paper is organized as follows. 
Section \ref{sec:literature_review} reviews the relevant literature. 
Section \ref{sec:methodology} describes the problem and the optimization model for TCMUM-AMoD, and proposes the first-order approximation algorithm for solving the problem at scale. 
Section \ref{sec:results} outlines experimental setups, data preparation, and displays experiment results and sensitivity analyses. 
Finally, Section \ref{sec:conclusion} recaps the main contributions of this work, outlines the limitations, and provides future research directions.

\section{Literature Review}
\label{sec:literature_review}

\subsection{Transit network design problem}

The Transit Network Design Problem (TNDP) is first proposed by \citet{CEDER1986331}, which can be separated into sub-problems ranging across tactical, strategical, and operational decisions, including Transit Network Design (TND), Frequency Setting (FS), Transit Network Timetabling (TNT), Vehicle Scheduling Problem (VSP), and Driver Scheduling Problem (DSP). Thorough reviews of TNDP and its sub-problems can be found in \citet{Ceder2007} and \citet{Ibarra_2015}.

This paper addresses the challenge of transit network design by tackling the frequency setting problem. This involves determining the optimal number of trips for a specified set of transit lines to ensure a high level of service within a planning period. The Transit Frequency Setting Problem (TFSP) is first studied by \citet{Newell1971} using analytical models, where total passenger waiting time is minimized under a fixed passenger demand setting. \citet{Furth1981} modelled the TFSP with a non-linear program that maximized the overall social welfare considering responsive demand. A heuristic-based algorithm was proposed to solve the non-linear program. \citet{Verbas2013} extended the model proposed by \citet{Furth1981} by considering multiple service patterns when determining transit frequencies, where a service pattern corresponds to a unique sequence of stops that are served by transit vehicles. 

Recently, demand uncertainty, which is intensified by the remote work~\cite{NBERw28731, caros2023emerging, caros2023impacts}, has been considered when setting transit frequencies. \citet{Gkiotsalitis2021} studied the frequency setting problem for autonomous minibuses considering demand uncertainty. They utilized a traditional Stochastic Optimization (SO) method, Sample Average Approximation (SAA), to address the demand uncertainty. \citet{guo2022transit} utilized the Robust Optimization (RO) approach for solving the TFSP under a single transit line setting with demand uncertainty. A heuristic-based approach, namely Transit Downsizing (TD), was proposed to solve the large-scale real-world TFSP efficiently. 

\subsection{Mobility-on-Demand system}

% Ride-hailing system
Ride-hailing platforms such as Uber and Lyft provide Mobility-on-Demand (MoD) services to millions of users globally every day. For an extensive overview of the ride-hailing system, one can refer to the review conducted by \citet{Wang_Yang_2019}. Studies on the ride-hailing system consists of analyzing demand (customers)~\cite{Young2019, Wang2024}, examining supply (drivers)~\cite{Hall2018, Castillo2022, GUO2023104233}, developing market structures~\cite{Cohen_Zhang, zhang2022economies, wang2023quantifying, GUO2023104397}, and designing operational strategies for platforms. The operational strategies typically include dynamic pricing~\cite{Banerjee2022, LIU2023103960}, customer-driver matching~\cite{Alonso-Mora462, Bertsimas2019, Tafreshian2020}, and vehicle rebalancing~\cite{Wen2017, GUO2021, Guo2022, guo2023fairnessenhancing}.

% AMoD system
Moreover, advancements in autonomous driving technology have introduced a new paradigm in transportation: Autonomous Mobility-on-Demand (AMoD)~\cite{Zardini_Lanzetti_Pavone_Frazzoli_2021}. The AMoD system has the potential to increase driver supply within the ride-hailing system while reducing service costs. Furthermore, the complete compliance of AMoD vehicles eliminates scenarios of driver rejections and leads to more efficient vehicle allocations. \citet{Iglesias_2017} introduced a Model Predictive Control (MPC) algorithm designed to optimizing rebalancing strategies, capitalizing on short-term demand forecasts through LSTM neural networks. Their proposed approach could significantly reduce the average customer wait time. \citet{Tsao2019} presented an MPC approach to optimize vehicle routes in AMoD system for both vacant and occupied vehicles. Their proposed algorithm has the potential to substantially decrease the distance traveled by mobility providers, thereby diminishing the impact of AMoD platforms on urban congestion. 

\subsection{Multimodal mobility system}

In recent years, a substantial body of research has illustrated the potential benefits of integrating MoD with traditional transit services to enhance urban mobility system.
\citet{Shen_Zhang_Zhao_2018} proposed an integrated AV-PT system where high-demand bus routes are maintained, low-demand bus routes are repurposed, and shared AVs are introduced as a complement for first-mile service during morning peak hours. An agent-based simulation was utilized to evaluate the integrated system performance. Their study revealed that the integrated system could potentially improve service quality, utilize road resources more efficiently, and being financially sustainable.
\citet{Wen_Chen_Nassir_Zhao_2018} proposed a systematic approach for the integration of AV-PT, concentrating on the development of AV solutions that complement and enhance existing transit networks. An agent-based simulation platform was developed to evaluate service performance, complemented by a discrete choice model of demand. Their results showed that the integrated system can significantly enhance urban mobility systems by improved service availability, reduced operational costs, and enhanced accessibility.
\citet{Mauro2020} studied the integration and coordination of AMoD systems with public transit to enhance urban mobility. They proposed a network flow model that maximized social welfare by optimizing the allocation of autonomous vehicles and their interaction with existing transportation infrastructure. Their results showed that integrating AMoD fleets with PT can yield considerable advantages, including improved mobility, reduced congestion, lower emissions, and enhanced system efficiency.

Other studies have focus on the network design and system operation of such an integrated system. \citet{Luo2021} proposed a framework for integrating micro-mobility services into transit network in order to connect packed urban centers to low-demand suburban areas, providing a low-cost, low-emission travel mode for short trips. They utilized a two-stage stochastic program to design the intermodal network considering demand uncertainty, with the first stage selecting transfer hub locations and the second stage optimizing system operations.

\citet{Steiner2020} developed a strategic network planning optimization model that incorporates MoD offering first-and-last-mile services. Their model simultaneously optimized bus line configurations, identifies zones for MoD service deployment, establishes MoD interactions with fixed-route networks through transfer points, and optimizes passenger routes based on specified service levels. However, the proposed model did not generate detailed transit schedules, and passengers' route and mode choices are not fully modelled, where only route choice in the bus network is considered.

\citet{Pinto2020} proposed a bi-level mathematical programming model for a joint system design problem with multimodal transit and shared AMoD. The upper level optimized transit network configurations and shared AMoD fleet sizes, while the lower level utilized an agent-based simulation to determine transit assignment and shared AMoD fleet operations with mode choice modeling. However, pricing for the integrated system is considered as an exogenous parameter.

\citet{banerjee2021realtime} explored the development of efficient routing policies for smart transit systems. These systems integrated high-capacity vehicles like buses with a fleet of cars to optimize the routing of trip requests in real-time, aiming to maximize social welfare within a specified time window. Nonetheless, passengers' mode and route choice behaviors were not considered and only a line configuration of bus networks was generated.

\citet{Luo2021B} addressed the joint optimization problem of transit network design and pricing for multimodal mobility systems. They aimed to determine optimal settings for mass transit frequencies, flows of MoD services, and pricing for each trip to maximize social welfare. The solution method included a primal-dual approach, a decomposition framework, and an approximation algorithm to solve optimization of large-scale problem instances. 

\citet{Wang2022} proposed an analytical framework for designing a transit-oriented multi-modal transportation system with passengers' route choices. They introduced a system-state equilibrium model that accounts for travelers' rational choice behaviors across different transportation modes and the corresponding impact on service levels. 
However, their study simplified the design of transit systems without generating a detailed network design and transit schedules.

\citet{Kumar2022} proposed a methodology framework for designing networks for an integrated system with MoD and transit. However, they did not allow shared rides, and transit lines are assumed to have unlimited capacity. Also, passengers' route and mode choice behavior is not considered. Their proposed method is demonstrated on small-sized networks (Sioux Falls).

\begin{table*}[h!]
    \caption{Research studies that solve the design of integrated MoD and PT system.}
    \label{tab:related_research}
    \centering
    \begin{adjustbox}{max width=\textwidth}
    \begin{tabular}{ l p{45mm} p{28mm} c p{25mm} p{35mm} p{35mm} }
     \hline 
     Paper & Transit decisions & MoD decisions & Pricing & Objectives & Demand modeling & Solution method \\
     \hline
     \citet{Luo2021} & Location of transfer hubs & Movement \newline of MoD vehicles & Yes & Maximize profit & Discrete \newline choice model & Two-stage stochastic program with heuristic algorithm\\
     \citet{Steiner2020} & Transit line configuration & MoD zones and transfer points & No & Minimize total \newline 
 system cost & Route assignment in transit network & MILP + branch-and-price algorithm\\
     \citet{Pinto2020} & Transit network design and frequency setting & Fleet sizing & No & Minimize travelers' disutility & Mode choice and route assignment & Bi-level programming + simulation\\
     \citet{banerjee2021realtime} & Transit line configuration and frequency setting & Movement \newline of MoD vehicles & No & Maximize \newline system welfare & No & MILP + Approximation algorithm \\
     \citet{Luo2021B} & Transit network design and frequency setting & Movement \newline of MoD vehicles & Yes & Maximize \newline system welfare & Route assignment & MILP + primal-dual approach + decomposition \\
     \citet{Wang2022} & Uniform stop distance and headway & Fleet sizing and allocation & Yes & Maximize \newline social welfare & No & System-state equilibrium model + search algorithm\\
     \citet{Kumar2022} & Transit line configuration and frequency setting & Fleet sizing and allocation & No & Minimize \newline travelers' cost & No & MILP + Benders decomposition \\
     This study & Transit network design and frequency setting & Fleet sizing and allocation & Yes & Minimize travelers' disutility & Discrete \newline choice model& MINLP + first-order approximation \\
     \hline
     \end{tabular}
     \end{adjustbox}
\end{table*}

Table \ref{tab:related_research} provides a summary of previous studies on the development of integrated MoD and PT systems. To the authors' knowledge, this work represents the first attempt to jointly solve network design and frequency setting for the transit system, fleet sizing and allocation for the MoD system, and pricing strategies for the combined system, all while taking into account passengers' mode and route choices. The proposed framework is motivated by \citet{Bertsimas_Sian_Yan_2020}, where a joint frequency-setting and pricing optimization problem is solved on a multimodal transit networks at scale. We further extend their methodology into the context of integrated PT and MoD system.

\section{Methodology}
\label{sec:methodology}

\subsection{Problem description}

Integrating the AMoD system into the existing transit network introduces a range of possible organizational structures. These structures can vary based on several factors, including the ownership of the transit and AMoD services, how these operators interact with each other, and the extent of regulation imposed by public authorities~\cite{SHEN2018125}. In this paper, we assume that the public transit agency is responsible for managing both transit services and AMoD operations. This setup involves the transit agency either owning the AMoD fleet outright or contracting with AMoD service providers. The purpose of operating such a transit-centric multimodal system for agencies is to leverage the AMoD fleet to enhance service delivery for passengers. This includes substituting low-demand bus lines with AMoD services and providing connections for passengers to and from rail stations.

With the existing transit network, our study focuses on the design of the TCMUM-AMoD system under the morning commute setting. Within this framework, we identify two distinct categories of commuters: \emph{local} commuters and \emph{downtown} commuters. Local commuters primarily rely on the transit system for short-distance trips within their local area, generally utilizing bus services. On the other hand, downtown commuters require services for longer-distance travel from suburban areas to downtown, typically facilitated through rail services or express bus services, to accommodate their commute needs efficiently. In the TCMUM-AMoD system, local commuters have two available mode options:
\begin{enumerate}
    \item \textbf{Bus}: local commuters take bus services to their destinations.
    \item \textbf{AMoD}: local commuters take AMoD services directly from their origins to their destinations.
\end{enumerate}
For downtown commuters, they have three available mode options:
\begin{enumerate}
    \item \textbf{Rail}: downtown commuters take rail services to their destinations.
    \item \textbf{Bus+Rail}: downtown commuters utilize local bus services to reach rail stations, from which they then board rail services to travel to their final destinations.
    \item \textbf{AMoD+Rail}: downtown commuters utilize AMoD services to reach rail stations, from which they then board rail services to travel to their final destinations.
\end{enumerate}
The rail services can also be replaced by express bus services for downtown commuters. Under this system setting, the AMoD only provides local trips to commuters and commuters have better accessibility to the transit network. 

% We consider the network design of the TCMUM-AMoD system under the morning commute setting where commuters have three available mode options: 
% \begin{itemize}
%     \item PT: commuters use the multimodal transit network including rail and bus services.
%     \item AMoD: commuters take a dedicated AMoD trip from their origins to their destinations.
%     \item PT-AMoD: commuters utilize both the multimodal transit network and the AMoD system to reach their destinations. The AMoD system offers both dedicated and shared FMLM trips connecting rail stations.
% \end{itemize}

In the design of the TCMUM-AMoD system, our approach involves the optimization of the rail and bus networks, the sizing and distribution of the local AMoD fleet, and the pricing structure for utilizing the AMoD system. The overarching goal of this optimization is to minimize the total disutility experienced by commuters within the system, which are quantified by waiting times and walking times. 

\begin{figure*}[!h]
\centering
\includegraphics[scale=0.6]{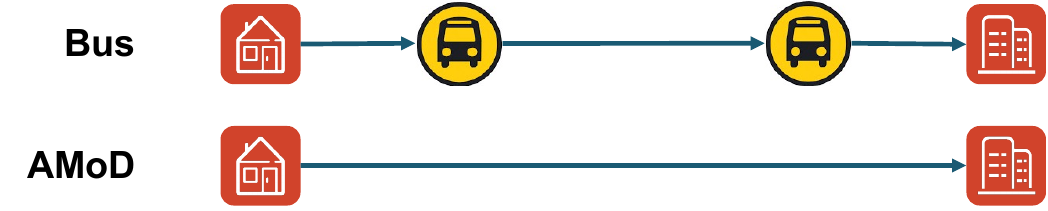}
\caption{Two route options for a morning local commute from home to the company. Commuters can: i) walk to the bus stop, take the bus and walk to the company, ii) take an AMoD service directly from home to the company.}
\label{fig:local_commuter}
\end{figure*}

\begin{figure*}[!h]
\centering
\includegraphics[scale=0.5]{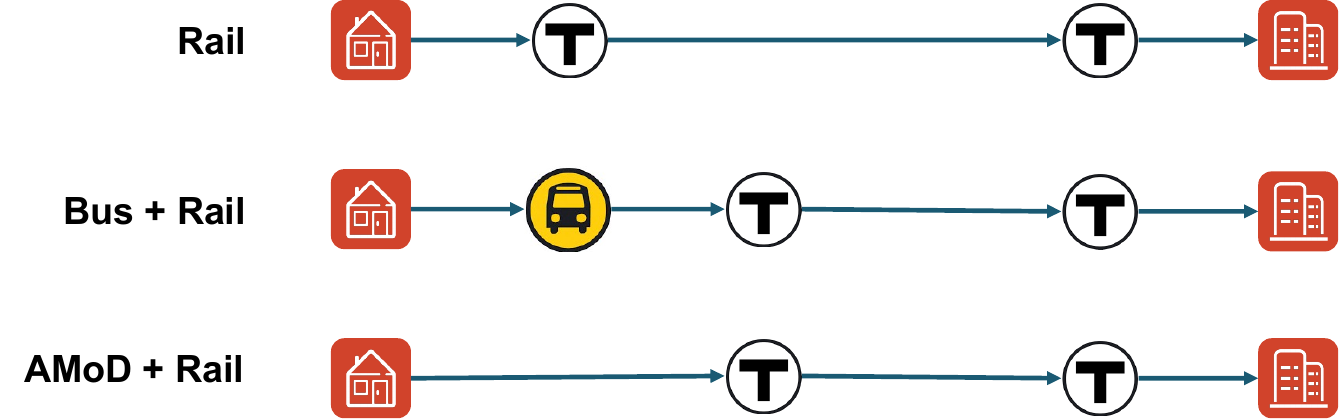}
\caption{Three route options for a morning downtown commute from home to the company. Commuters can: i) walk to the subway station, take the subway and walk to the company, ii) take a bus service to the subway station, take the subway and walk to the company, and iii) take an AMoD service to the subway station, take the subway and walk to the company.}
\label{fig:downtown_commuter}
\end{figure*}

Let $G = (V, E)$ denote the road network and a \textit{commute} in this problem is referred to as an origin-destination pair $(u,v)$, where $u,v \in V$. 
Let $\mathcal{U}$ indicate the set of commutes in the problem.
A commute $(u,v) \in 
\mathcal{U}$ could have multiple \textit{route} choices, denoted by the set $\mathcal{R}^{u,v}$, and each route corresponds to a distinct sequence of mode and path choices.
Each route $r \in \mathcal{R}^{u,v}$ contains at least one $\textit{leg}$, indicating a trip stage along path $r$. 
Let $\mathcal{J}(r)$ indicate the set of legs in route $r$.
Figure \ref{fig:local_commuter} and Figure \ref{fig:downtown_commuter} show instances with route options for a local commuter and a downtown commuter, respectively. %, which includes one single-leg route and three multi-leg routes.
% Meanwhile, if a route $r$ contains a transfer at a rail station, there will be two legs connected by a station in $r$.

The full study time period $[T_{start}, T_{end}]$ is divided into $T$ identical time intervals with length $\Delta_T$. Let $\mathcal{T} = \{1,2,...,T\}$ stand for the set of time intervals.
The commute demand data is indicated by the OD-matrix $\boldsymbol{d} = (d^{u,v}_t)$ where commute $(u,v)$ starts from origin $u$ go to destination $v$ at time $t$.
Given the commute demand at time $d^{u,v}_t$, commuters make route choices $\boldsymbol{\theta} = (\theta^{u,v,r}_t)$ among a set of routes, where $\theta^{u,v,r}_t$ represents the proportion of commuters for commute $(u,v)$ at time $t$ who chooses route $r$.

For the multimodal transit network design, let $\mathcal{L}^R$ and $\mathcal{L}^B$ denote the set of rail and bus $\textit{lines}$, where a line indicates a sequence of stops.
Let $\mathcal{L}$ represent the set of transit lines, i.e., $\mathcal{L} =  \mathcal{L}^R \cup \mathcal{L}^B$.
For any transit line $l \in \mathcal{L}$, the decision variable is $\boldsymbol{x} = (x_t^l)$, where $x_t^l \in \mathbb{Z}$ indicates the number of departures from the start of the line $l$ at time $t$. 
A lower bound $\underline{B}_R$ and an upper bound $\overline{B}_R$ are imposed on the number of departures of rail lines to guarantee the minimum level of service and the minimum headway between two consecutive trains.
Similarly, an upper bound $\overline{B}_B$ is enforced to the number of departures of bus lines for the purpose of maintaining the minimum headway condition. 
Setting decision variable $x_t^l$ equal to 0 is equivalent to removing the bus line $l$ from the transit network at time $t$.
Given the total budget $B_{rail}$ and $B_{bus}$ for the transit network, which denotes the total number of rail and bus services that can be offered, the set of feasible transit networks is denoted by
\begin{equation} \small
    \mathcal{X} = \left\{ (\boldsymbol{x}^R, \boldsymbol{x}^B) \in \mathbb{Z}_{+}^{T \times |\mathcal{L}|}: \sum_{t \in \mathcal{T}} \sum_{l \in \mathcal{L}^B} c^l x_t^l \leq B_{bus}, \sum_{t \in \mathcal{T}} \sum_{l \in \mathcal{L}^R} c^l x_t^l \leq B_{rail}, \underline{B}_R \leq \boldsymbol{x}^R \leq \overline{B}_R, \boldsymbol{x}^B \leq \overline{B}_B \right\},
\end{equation}
where $\boldsymbol{x}^R$ and $\boldsymbol{x}^B$ are vectors of decision variables for rail and bus lines, and $c^l$ stands for the cost for line $l$, which varies across different transit lines. Instead of solving for a detailed transit schedule, we relax the feasibility constraint (1) to generate a set of departure rates of transit services during each time interval:
\begin{equation} \small
   \Bar{\mathcal{X}} = \left\{ (\boldsymbol{x}^R, \boldsymbol{x}^B) \in \mathbb{R}_+^{T \times |\mathcal{L}|}: \sum_{t \in \mathcal{T}} \sum_{l \in \mathcal{L}^B} c^l x_t^l \leq B_{bus}, \sum_{t \in \mathcal{T}} \sum_{l \in \mathcal{L}^R} c^l x_t^l \leq B_{rail}, \underline{B} \leq \boldsymbol{x}^R \leq \overline{B}, \boldsymbol{x}^B \leq \overline{B} \right\}.
\end{equation}
The relaxed feasible schedules $\Bar{\mathcal{X}}$ are straightforward to interpret and more flexible for transit operators to follow. For instance, $x_t^l = 1.5$ indicates that line $l$ departs twice every three time intervals.
Also, the relaxed feasibility set only requires solving a Linear Programming (LP) problem.

The AMoD system offers both \emph{local} and \emph{first-mile-last-mile (FMLM)} trips to commuters. The local AMoD trip indicates that local commuters take an AMoD trip directly from their origins to their destinations.
The FMLM trips connect downtown commuters' with rail stations.
In this paper, we assume the existence of a fleet of vehicles tasked with providing both direct and FMLM AMoD services. Our primary focus is on optimizing the sizing and allocation of this vehicle fleet and determining the pricing strategies for using these services.

Let $\mathcal{S}$ denote the set of rail stations in the transit system. 
The decision variable for the fleet size of AMoD vehicles is $\boldsymbol{N} = (N_t^s)$, where $N_t^s \in \mathbb{Z}_+$ indicates the number of AMoD vehicles within the \textit{nearby} region of rail station $s \in S$ at time $t$.
The nearby region of each rail station $s$ is predefined and has area $A_s$.
Given the maximum number of AMoD vehicles $\overline{N}$, the set of feasible AMoD vehicle allocations is indicated by
\begin{equation} \small
\label{eq:N}
    \mathcal{N} = \left \{  \boldsymbol{N} \in \mathbb{Z}_+^{|\mathcal{S}|}: \sum_{s \in \mathcal{S}} N_t^s \leq \overline{N}, \forall t \in \mathcal{T} \right \}. 
\end{equation}
We further relax the feasible vehicle allocation set $\mathcal{N}$ to reduce the computation complexity:
\begin{equation} \small
\label{eq:N}
    \Bar{\mathcal{N}} = \left \{  \boldsymbol{N} \in \mathbb{R}_+^{|\mathcal{S}|}: \sum_{s \in \mathcal{S}} N_t^s \leq \overline{N}, \forall t \in \mathcal{T} \right \}. 
\end{equation}
The feasible allocation of vehicles around rail station $s$ at time $t$ can be obtained by rounding down $N_t^s \in \Bar{\mathcal{N}}$.

\subsection{Optimization with known path choices and non-shared local AMoD fleet}

First, we formulate the optimization problem with known path choices $\boldsymbol{\theta}$ (satisfying $0 \leq \theta_t^{u,v,r} \leq 1$ and $\sum_{r \in \mathcal{R}^{u,v}} \theta_t^{u,v,r} = 1, \forall t \in 
\mathcal{T}$) and an AMoD fleet which only offers non-shared services to commuters. Both assumptions will be revisited and relaxed in the following sections. 

Let $(u,v,r)$ denote a \emph{commute route} for route $r$ which serves commute $(u,v)$, and $(u,v,r,i)$ indicate a \emph{commute-route leg} corresponds to the $i$-th leg of the itinerary of commute route $(u,v,r)$.
Let $\boldsymbol{z} = (z_t^{u,v,r,i})$ denote the boarding variables for the commute-route leg $(u,v,r,i)$, where $z_t^{u,v,r,i} \in \mathbb{R}_+$ indicates the number of commuters traveling from $u$ to $v$, choosing route $r$, on the $i$-th leg of their itinerary, boarding a transit (or AMoD) service that starts (or is available) at time $t$.
For the simplicity of the model formulation, we assume that the travel time on any trip leg is zero\footnote{Incorporating the travel time for trip legs only requires adjusting time indices in the model, which significantly complicates the model formulation. It is straightforward to incorporate travel times in experiments.}.
The system design of the TCMUM-AMoD system can be formulated as
\begin{equation}
\label{eq:static_problem}
    \min_{\boldsymbol{x} \in \Bar{\mathcal{X}}, \boldsymbol{N} \in \Bar{\mathcal{N}}} Q(\boldsymbol{x}, \boldsymbol{N}, \boldsymbol{\theta}),
\end{equation}
where 
\begin{subequations}
\small
\label{eq:base_problem}
\begin{align}
    Q(\boldsymbol{x}, \boldsymbol{N}, \boldsymbol{\theta}) = \; & \min_{\boldsymbol{z} \geq 0} \; && J_{Transit}(\boldsymbol{z}, \boldsymbol{\theta}, \boldsymbol{x}) + J_{AMoD}(\boldsymbol{z}, \boldsymbol{\theta}, \boldsymbol{N}) \\
    &\text{s.t.} \; 
    && O_{l,s,t}(\boldsymbol{z}) \leq K^l x_t^l  &&&\forall l \in \mathcal{L}, \forall s \in \mathcal{S}_l, \forall t \in \mathcal{T}; \\
    & && P_{s,t}(\boldsymbol{z}) \leq \frac{\Delta_T}{\mathbb{E}[T_l^s]} N_t^s  &&&\forall s \in \mathcal{S}, \forall t \in \mathcal{T}; \\
    & && \sum_{t = 1}^\tau z_{t}^{u,v,r,1} \leq \sum_{t = 1}^\tau d^{u,v}_t \theta^{u,v,r}_{t} &&&\forall u,v \in \mathcal{U}, \forall r \in \mathcal{R}^{u,v}, \forall \tau \in \mathcal{T}; \\
    & && \sum_{t = 1}^\tau z_{t}^{u,v,r,i} \leq \sum_{t = 1}^\tau z_{t}^{u,v,r,i-1}  &&&\forall u,v \in \mathcal{U}, \forall r \in \mathcal{R}^{u,v}, \forall i = 2,..., |\mathcal{J}(r)|, \forall \tau \in \mathcal{T}.
\end{align}
\end{subequations}
The functions $J_{Transit}(\boldsymbol{z}, \boldsymbol{\theta}, \boldsymbol{x})$ and $J_{AMoD}(\boldsymbol{z}, \boldsymbol{\theta}, \boldsymbol{N})$ in Equation (\ref{eq:base_problem}a) indicate the total waiting and walking time of the TCMUM-AMoD system. In our model, the waiting time for commuters in transit systems consist of both the expected waiting time given the line frequency and the excess waiting time. The expected waiting time for a commute route $(u, v, r)$ traveling at time $t$ can be calculated as
$$\sum_{i \in \mathcal{A}(u,v,r)} z_t^{u,v,r,i} \cdot \frac{\Delta_T}{2 \cdot x_t^{l(i)}},$$
where $\mathcal{A}(u,v,r)$ indicates the set of transit legs in the commute route $(u,v,r)$ and $l(i)$ represents the transit line using by the transit leg $i \in \mathcal{A}(u,v,r)$. The total expected waiting time for the transit system is
\begin{equation}
\label{eq:total_transit_expected_wait_time}
    J_{Transit}^{exp-wait}(\boldsymbol{z}, \boldsymbol{x}) = \sum_{(u,v) \in \mathcal{U}} \sum_{r \in \mathcal{R}^{u,v}} \sum_{t \in \mathcal{T}} \sum_{i \in \mathcal{A}(u,v,r)} z_t^{u,v,r,i} \cdot \frac{\Delta_T}{2 \cdot x_t^{l(i)}}.
\end{equation}
The Equation (\ref{eq:total_transit_expected_wait_time}) is ill-defined as $x_t^{l(i)}$ could be zeros. However, Constraints (\ref{eq:base_problem}b) guarantee that $z_t^{u,v,r,i}$ will be zero if $x_t^{l(i)}$ is zero. Therefore, we assume $z_t^{u,v,r,i} \cdot \frac{\Delta_T}{2 \cdot x_t^{l(i)}} = 0$ whenever $x_t^{l(i)}=0$.

% In our model, the waiting time for commuters is defined as the additional time intervals they must wait before boarding the next available service. Specifically, if a commuter arrives at a bus stop within a given time interval and a bus also arrives within the same time interval, it is assumed that the commuter will not experience any waiting time\footnote{There will be waiting time in this scenario, including the detailed waiting time in the objective function will lead to a non-linear problem which is challenging to solve. The current definition of waiting time provides a reasonable approximation of total waiting time when commuters' route choice are pre-determined or modelled accurately.}. 

For the excess waiting time, the total number of commuters waiting at a station (or stop) $s$ on line $l$ at time $\tau$ can be computed by
\begin{equation}
\label{eq:pt_wait_time}
    W_{Transit}^{s,l,\tau}(\boldsymbol{z}, \boldsymbol{\theta}) = AD^{s,l,\tau}(\boldsymbol{\theta}) + XD^{s,l,\tau}(\boldsymbol{z}) - BD^{s,l,\tau}(\boldsymbol{z}),
    %\sum_{(u,v,r,i) \in \mathcal{H}(s,l)} \sum_{t=1}^\tau \left ( z_t^{u,v,r,i-1} - z_t^{u,v,r,i} \right ),
\end{equation}
where $AD^{s,l,\tau}(\boldsymbol{\theta})$ indicates ``\emph{arrival demand}'', $XD^{s,l,\tau}(\boldsymbol{z})$ represents ``\emph{transferring demand}'', and $BD^{s,l,\tau}(\boldsymbol{z})$ corresponds to ``\emph{boarding demand}'', all cumulative up until time $\tau$. $W_{Transit}^{s,l,\tau} (\boldsymbol{z}, \boldsymbol{\theta})$ represents the number of commuters who are at station $s$ but not able to board line $l$ at time $\tau$. Then the total excess waiting time for the transit system can be obtained by aggregating Equation (\ref{eq:pt_wait_time}) as
\begin{equation}
\label{eq:total_excess_transit_wait_time}
    J_{Transit}^{exc-wait}(\boldsymbol{z}, \boldsymbol{\theta}) = \sum_{l \in \mathcal{L}} \sum_{s \in \mathcal{S}_l} \sum_{\tau \in \mathcal{T}} \left[ AD^{s,l,\tau}(\boldsymbol{\theta}) + XD^{s,l,\tau}(\boldsymbol{z}) - BD^{s,l,\tau}(\boldsymbol{z}) \right] \cdot \Delta_T,
\end{equation}
where $\mathcal{S}_l$ represents the set of stations for line $l$. 

The arrival demand $AD^{s,l,\tau}(\boldsymbol{\theta})$ in Equations (\ref{eq:pt_wait_time}) and (\ref{eq:total_excess_transit_wait_time}) indicates the total arrival demand that has arrived at station $s$ on line $l$ by time $\tau$ and it is computed by 
\begin{equation}
\label{eq:arrival_demand}
    AD^{s,l,\tau}(\boldsymbol{\theta}) = \sum_{(u,v,r) \in \mathcal{K}(s, l)} \sum_{t=1}^{\tau} d_t^{u, v} \theta_t^{u, v, r},
\end{equation}
where the set $\mathcal{K}(s, l)$ denotes the set of commute routes that board on station $s$ on line $l$ first. 

The transferring demand $XD^{s,l,\tau}(\boldsymbol{z})$ represents the total number of passengers who have arrived at station $s$ on line $l$ by time $\tau$, having transferred over from another transit line. It can be formulated as
\begin{equation}
\label{eq:transfer_demand}
    XD^{s,l,\tau}(\boldsymbol{z}) = \sum_{(u,v,r,i) \in \mathcal{H}(s, l)} \sum_{t=1}^{\tau} z_t^{u,v,r,i-1},
\end{equation}
where the set $\mathcal{H}(s,l)$ denotes the set of commute-route legs that make a transfer through station $s$ on line $l$. For a commute-route leg $(u,v,r,i)$ to be considered in the set $\mathcal{H}(s,l)$, it must satisfy the following criteria: i) $i \geq 2$, ii) the transfer station $s$ connects the $(i-1)$-th and $i$-th legs of the commute route $(u, v, r)$, and iii) commute-route leg $i$ utilizes the transit line $l$.

The boarding demand $BD^{s,l,\tau}(\boldsymbol{z})$ indicates the total number of passengers who have managed to board a transit vehicle at station $s$ on line $l$ by time $\tau$, which can be formulated as
\begin{equation}
\label{eq:boarding_demand}
    BD^{s,l,\tau}(\boldsymbol{z}) = \sum_{(u,v,r,i) \in \mathcal{U}(s, l)} \sum_{t=1}^{\tau} z_t^{u,v,r,i},
\end{equation}
where the set $\mathcal{U}(s, l)$ represents all of the commute-route legs that require boarding a transit vehicle at station $s$ on line $l$. This includes commute routes that start from station $s$ on line $l$ or later transfer to line $l$ through station $s$.

For the walking time of commuters in transit systems, it can be formulated as
\begin{equation}
\label{eq:total_walking_time_transit}
    J_{Transit}^{walk}(\boldsymbol{z}) = \sum_{(u,v) \in \mathcal{U}} \sum_{r \in \mathcal{R}^{u,v}} \sum_{t \in \mathcal{T}} z_t^{u,v,r,1} \cdot w(u,v,r),
\end{equation}
where $w(u,v,r)$ denotes the walking time for a commute route $(u, v, r)$. And the total disutility of using the transit system is
\begin{equation}
\label{eq:total_waiting_time_transit}
    J_{Transit}(\boldsymbol{z}, \boldsymbol{\theta}, \boldsymbol{x}) = J_{Transit}^{exp-wait}(\boldsymbol{z}, \boldsymbol{x}) + J_{Transit}^{exc-wait}(\boldsymbol{z}, \boldsymbol{\theta}) + J_{Transit}^{walk}(\boldsymbol{z}).
\end{equation}

The AMoD system provides commuters with local or FMLM services. The AMoD system provides door-to-door services to commuters, therefore, commuters do not experience walking time when using AMoD services. For the waiting time, it consists of the expected waiting time given the number of AMoD vehicles nearby and the excess waiting time. The expected waiting time for a commute route $(u, v, r)$ traveling at time $t$ can be calculated as
$$\sum_{i \in \mathcal{B}(u,v,r)} z_t^{u,v,r,i} \cdot \frac{\alpha_s}{\Bar{v}} \sqrt{A_s / N_t^s},$$
where $\mathcal{B}(u,v,r)$ indicates the set of AMoD legs in the commute route $(u, v, r)$, and $\frac{\alpha_s}{\Bar{v}} \sqrt{A_s / N_t^s}$ is the expected wait time for AMoD services around station $s$~\citep{Urban_OR}. $\Bar{v}$ indicates the average local AMoD vehicle speed and $\alpha_s$ is a parameter depending on the shape of nearby region and the location of station $s$. The total expected waiting time for the AMoD system is
\begin{equation}
\label{eq:AMoD_waiting_time_expected}
    J_{AMoD}^{exp-wait}(\boldsymbol{z}, \boldsymbol{N}) = \sum_{(u,v) \in \mathcal{U}} \sum_{r \in \mathcal{R}^{u,v}} \sum_{t \in \mathcal{T}} \sum_{i \in \mathcal{B}(u,v,r)} z_t^{u,v,r,i} \cdot \frac{\alpha_s}{\Bar{v}} \sqrt{A_s / N_t^s}.
\end{equation}

Similarly, Equation (\ref{eq:AMoD_waiting_time_expected}) is ill-defined as $N_t^s$ could be zero. Constraints (\ref{eq:base_problem}c) guarantee that $z_t^{u,v,r,i}$ will be zero if $N_t^s$ is zero. Therefore, we assume the term $z_t^{u,v,r,i} \cdot \frac{\alpha_s}{\Bar{v}} \sqrt{A_s / N_t^s} = 0$ whenever $N_t^s = 0$.

For the excess waiting time of the AMoD system, the local AMoD trips originating near station $s$, the number of local commuters waiting to get AMoD trips at time $\tau$ can be formulated as 
\begin{equation}
    W_{Direct}^{s,\tau} (\boldsymbol{z}, \boldsymbol{\theta}) = \sum_{(u,v,r) \in \mathcal{Y}(s)} \sum_{t = 1}^\tau \left ( d_t^{u,v} \theta_t^{u,v,r} - z_t^{u,v,r,1} \right ),
\end{equation}
where the set $\mathcal{Y}(s)$ denotes the set of local commutes that begin in close proximity to station $s$ and utilize AMoD vehicles to reach their destinations directly.
Similarly, for the first-mile trips, the number of downtown commuters waiting to get first-mile AMoD services to a rail station $s$ at time $\tau$ can be computed by
\begin{equation}
    W_{First}^{s,\tau} (\boldsymbol{z}, \boldsymbol{\theta}) = \sum_{(u,v,r) \in \mathcal{M}(s)} \sum_{t = 1}^\tau \left ( d_t^{u,v} \theta_t^{u,v,r} - z_t^{u,v,r,1} \right ),
\end{equation}
where the set $\mathcal{M}(s)$ indicates the set of downtown commute routes that take AMoD services from their origins to the rail station $s$. The number of downtown commuters waiting at a rail station $s$ at time $\tau$ to get last-mile AMoD services to their destinations can be formulated as
\begin{equation}
     W_{Last}^{s,\tau} (\boldsymbol{z}) = \sum_{(u,v,r,i) \in \mathcal{N}(s)} \sum_{t = 1}^\tau \left ( z_t^{u,v,r,i-1} -  z_t^{u,v,r,i} \right ),
\end{equation}
where the set $\mathcal{N}(s)$ denotes the set of commute-route legs that take AMoD services from station $s$ to their destinations. Overall, the total excess waiting time for the AMoD system is
\begin{equation}
    J_{AMoD}^{exc-wait}(\boldsymbol{z}, \boldsymbol{\theta}) = \sum_{s \in \mathcal{S}} \sum_{\tau \in \mathcal{T}} \left [ W_{Direct}^{s,\tau} (\boldsymbol{z}, \boldsymbol{\theta}) + W_{First}^{s,\tau} (\boldsymbol{z}, \boldsymbol{\theta}) + W_{Last}^{s,\tau} (\boldsymbol{z}) \right ] \cdot 
    \Delta_T,
\end{equation}
and the total disutility for the AMoD system is 
\begin{equation}
\label{eq:total_waiting_time_AMoD}
    J_{AMoD}(\boldsymbol{z}, \boldsymbol{\theta}, \boldsymbol{N}) = J_{AMoD}^{exp-wait}(\boldsymbol{z}, \boldsymbol{N}) + J_{AMoD}^{exc-wait}(\boldsymbol{z}, \boldsymbol{\theta}).
\end{equation}

Constraints (\ref{eq:base_problem}b) guarantee that the number of commuters on board a transit vehicle does not exceed the capacity $K^l$. $O_{l,s,t}(\boldsymbol{z})$ represents the occupancy of a transit vehicle, which starts to operate at time $t$, as it passes the station (or stop) $s$ on line $l$. It is formulated as
\begin{equation}
    O_{l,s,t}(\boldsymbol{z}) = \sum_{(u,v,r,i) \in \mathcal{I}(s,l)} z_t^{u,v,r,i},
\end{equation}
where the set $\mathcal{I}(s,l)$ incorporates commute-route legs that pass through the station $s$ on line $l$. For a commute-route leg $(u,v,r,i)$ to be considered in the set $\mathcal{I}(s,l)$, it must satisfy the following criteria: i) the $i$-th leg utilizes the transit line $l$, ii) the transfer station connecting the $(i-1)$-th and $i$-th legs is \emph{at or before} station $s$ on line $l$, and iii) the transfer station connecting $i$-th and $(i+1)$-th legs is \emph{after} station $s$ on line $l$. 

Constraints (\ref{eq:base_problem}c) ensure that the number of vehicles providing AMoD services does not exceed the number of available AMoD vehicles $\frac{\Delta_T}{\mathbb{E}[T_l^s]} N_t^s$ in the nearby region of station $s$ at time $t$. Given that not every AMoD vehicle will be accessible to commuters at each time interval due to some vehicles being occupied with serving existing demand, we employ the expression $\frac{\Delta_T}{\mathbb{E}[T_l^s]}$ to approximate the average availability rate of AMoD vehicles near the station $s$. The ratio is constructed as follows: i) let the average local trip distance with both origins and destinations within the nearby region of station $s$ be $\mathbb{E}[D_l^s] = \alpha_s \sqrt{A_s}$, while the value of $\alpha_s$ depends on the shape of nearby region and the location of station $s$~\cite{Urban_OR}; ii) assume an average local AMoD vehicle speed of $\Bar{v}$; iii) the average local AMoD trip time is $\mathbb{E}[T_l^s] = \frac{\mathbb{E}[D_l^s]}{\Bar{v}}$; iv) the average availability rate of AMoD vehicles near the station $s$ can then be formulated as $\frac{\Delta_T}{\mathbb{E}[T_l^s]}$.

$P_{s,t}(\boldsymbol{z})$ denotes the number of AMoD trips near the rail station $s$ at time $t$, and it can be computed by
\begin{equation}
    P_{s,t}(\boldsymbol{z}) = \sum_{(u,v,r) \in \mathcal{Y}(s) \bigcup \mathcal{M}(s)} z_t^{u,v,r,1} + \sum_{(u,v,r,i) \in \mathcal{N}(s)} z_t^{u,v,r,i}.
\end{equation}
Constraints (\ref{eq:base_problem}d) make sure that the number of commuters boarding the first leg of a commute route $(u,v,r)$ up until time $\tau$ should not exceed the total demand for the commute route $(u,v,r)$ up until time $\tau$. Constraints (\ref{eq:base_problem}e) impose that commuters with a commute route $(u, v, r)$ can board $i$-th trip leg only if they complete $(i-1)$-th trip leg in the itinerary.

Both functions $J_{Transit}(\boldsymbol{z}, \boldsymbol{\theta}, \boldsymbol{x})$ and $J_{AMoD}(\boldsymbol{z}, \boldsymbol{\theta}, \boldsymbol{N})$ are nonlinear with respect to decision variables $\boldsymbol{x}$ and $\boldsymbol{N}$. Overall, the problem (\ref{eq:static_problem}) is a nonlinear program that is intractable for large-scale instances. We will introduce a heuristic to linearize and solve the problem later.

\subsection{Serving commuters with shared AMoD fleet}

\begin{figure}[!h]
    \centering
    \begin{subfigure}[b]{0.4\textwidth}
        \centering
        \includegraphics[width=\textwidth]{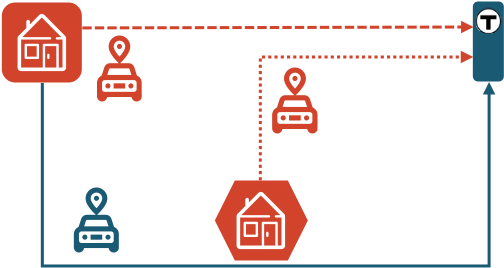}
        \caption[]%
        {First-mile AMoD scenarios}
        \label{}
    \end{subfigure}
    \hfill
    \begin{subfigure}[b]{0.58\textwidth}  
        \centering 
        \includegraphics[width=\textwidth]{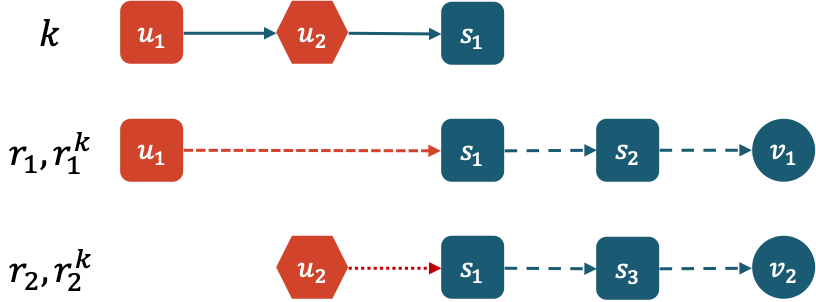}
        \caption[]%
        {Related commuter routes and a shared first-mile AMoD trip}
        \label{}
    \end{subfigure}
    \caption{Example explaining the route separation after introducing shared AMoD services. Trip $k$ indicates a first-mile shared AMoD trip which is shared by commutes $(u_1,v_1)$ and $(u_2, v_2)$. Each commute contains both routes with non-shared AMoD trips $r_1, r_2$ and routes with shared AMoD trips $r_1^k, r_2^k$. Routes $r_1$ ($r_2$) and $r_1^k$ ($r_2^k$) share the same itinerary but different first-mile AMoD services.} 
    \label{fig:shared_AMOD_example}
\end{figure}

In this section, we consider scenarios with a shared AMoD fleet. After incorporating shared AMoD services into the problem, the same commute route including AMoD trips should be separated into multiple commute routes indicating different sharing scenarios. Figure \ref{fig:shared_AMOD_example} provides an instance for explaining the route separation.

% For instance, if a commute route $(u,v,r)$ includes a first-mile AMoD trip and the commute $(u,v)$ can share the first-mile AMoD trip with another commute $(u', v')$, there exist two routes $r, r'$ with the same itinerary where $r$ includes a dedicated AMoD trip and $r'$ contains a shared AMoD trip. Figure \ref{fig:shared_AMOD_example} provides an instance for explaining the route separation.

Let $\mathcal{P}$ denote the set of shared AMoD trips.
For a shared trip $p \in \mathcal{P}$, let $\mathcal{R}(p)$ represent the set of commute routes that incorporates the shared trip $p$. Meanwhile, let $\mathcal{Q}(p)$ indicate the set of commutes that get involved in the shared trip $p$. For a commute $(u,v) \in \mathcal{Q}(p)$, let $\mathcal{R}^{u,v}(p)$ denote the set of routes for the commute $(u,v)$ that includes the shared trip $p$. Then we introduce additional constraints to the optimization problem:
\begin{equation}
\label{eq:share_conservation}
    \sum_{r \in \mathcal{R}^{u,v}(p)} z_t^{u,v,r,i} = \sum_{r' \in \mathcal{R}^{u',v'}(p)} z_t^{u',v',r',i'} \quad \forall p \in \mathcal{P}, \forall (u,v), (u',v') \in \mathcal{Q}(p), \forall t \in \mathcal{T},
\end{equation}
where $i$ and $i'$ are indices of commute-route legs in routes $r$ and $r'$ that correspond to the shared trip $p$, respectively.
Constraint (\ref{eq:share_conservation}) maintains the consistency of boarding variables among commute-route legs that correspond to the same shared trip.

When allowing shared AMoD services, the number of AMoD trips does not equal to the number of vehicles utilized for providing AMoD services. 
Therefore, we define $\Tilde{P}_{s,t}(\boldsymbol{z})$ as the number of vehicles used for providing AMoD trips around rail station $s$ at time $t$ with sharing, which is formulated as
\begin{equation}
    \Tilde{P}_{s,t}(\boldsymbol{z}) = \sum_{(u,v,r) \in \mathcal{Y}(s) \bigcup \mathcal{M}(s)} \xi^{u,v,r,1} z_t^{u,v,r,1} + \sum_{(u,v,r,i) \in \mathcal{N}(s)} \xi^{u,v,r,i} z_t^{u,v,r,i},
\end{equation}
where $\xi^{u,v,r,i}$ indicates the vehicle discount factor for a commute-route leg $(u,v,r,i)$ which corresponds to a shared AMoD trip. The value of $\xi^{u,v,r,i}$ depends on the number of commutes sharing the same AMoD trip. For instance, $\xi^{u,v,r,i} = 0.5$ if the commute-route leg $i$ corresponds to a shared AMoD trip with another commute $(u',v')$. In general, $\xi^{u,v,r,i} = \frac{1}{n}$ where $n$ is the number of commutes included in the shared AMoD trip. 

The optimization problem with shared AMoD fleet can then be formulated as
\begin{equation}
    \min_{\boldsymbol{x} \in \Bar{\mathcal{X}}, \boldsymbol{N} \in \Bar{\mathcal{N}}} \Tilde{Q}(\boldsymbol{x}, \boldsymbol{N}, \boldsymbol{\theta}),
\end{equation}
where 
\begin{subequations}
\label{eq:base_problem_shared}
\begin{align}
    \Tilde{Q}(\boldsymbol{x}, \boldsymbol{N}, \boldsymbol{\theta}) = &  \min_{\boldsymbol{z} \geq 0}  && J_{Transit}(\boldsymbol{z}, \boldsymbol{\theta}, \boldsymbol{x}) + J_{AMoD}(\boldsymbol{z}, \boldsymbol{\theta}, \boldsymbol{N}) \\
    &\text{s.t.} \; 
    && \text{Constraints} \; (\ref{eq:base_problem}b), (\ref{eq:base_problem}d), (\ref{eq:base_problem}e), (\ref{eq:share_conservation}),\\
    & &&  \Tilde{P}_{s,t}(\boldsymbol{z}) \leq \frac{\Delta_T}{\mathbb{E}[T_l^s]} N_t^s  &&&\forall s \in \mathcal{S}, \forall t \in \mathcal{T}.
\end{align}
\end{subequations}

\subsection{Optimization with design-dependent choices}

In this section, we introduce a framework that not only facilitates the generation of the optimal design for a multimodal system but also enables the service operator to adjust pricing structures, denoted as $\boldsymbol{p}(\lambda) = (p^{u,v,r}(\lambda))$. Here, $p^{u,v,r}(\lambda)$ represents the price levied for a commute between points $u$ and $v$ via route $r$. This paper primarily concentrates on the modification of a discount factor $\lambda \in [\underline{\lambda}, \Bar{\lambda}]$, which is utilized for the AMoD services. Regarding the pricing of transit services and baseline AMoD services, it is presumed that their fares are predetermined and treated as exogenous parameters within our analysis.

For transit services, we assume a flat fare $f^l$ for each transit line $l \in \mathcal{L}$ and a discount factor $\nu$ for transfers within the multimodal transit system. In practice, a transfer from the rail system to the bus system is typically free within a time period, i.e., $\nu = 0$ for taking the bus line.

For the baseline AMoD fare, we assume it follows the standard fare structure of Transportation Network Companies (TNCs), e.g., Uber or Lyft. The fare structure consists of a base fare $f^{base}$, a book fare $f^{book}$, a minimum fare $f^{min}$, a distance rate $\pi_{d}$ and a time rate $\pi_{t}$. Given an AMoD trip with distance $d$ and travel time $\tau$, the price $p$ will be
$$p^{AMoD} (d, \tau) = \max(f^{base} + f^{book} + \pi_{d} \cdot d + \pi_{t} \cdot \tau, f^{min}).$$

For a route $r \in \mathcal{R}^{u,v}$ of a commute $(u,v)$, let $\mathcal{J}^{transit}(r)$ and $\mathcal{J}^{AMoD}(r)$ denote the set of legs in $r$ which corresponds to commuters taking transit lines and AMoD vehicles, respectively. 
For each leg $j \in \mathcal{J}(r)$, let $d_j$ represent the travel distance and $\tau_j$ indicate the travel time. 
For a transit leg $j \in \mathcal{J}^{transit}(r)$, let $l(j)$ stand for the transit line corresponding to the leg $j$. Let $\hat{l}(r)$ represent the first transit line in route $r$ if the route $r$ does not contain any AMoD legs. For a route $r$ including AMoD legs, $\hat{l}(r) = \emptyset$ is an empty set.
The price $p^{u,v,r}$ can then be formulated as
\begin{equation} 
\label{eq:prices}
\small
    p^{u,v,r} (\lambda) = f^{\hat{l}(r)} + \sum_{j \in \mathcal{J}^{transit}(r) \setminus \{\hat{l}(r)\}} \nu f^{l(j)}  + \sum_{j \in \mathcal{J}^{AMoD}(r)} \lambda \cdot p^{AMoD}(d_j, \tau_j).
\end{equation}
It is worth noting that the pricing structures for both transit and AMoD operators can be adjusted seamlessly without changing the overall optimization model. Future extensions with more complicated and practical pricing structures can be considered. 

By allowing the design-dependent choices for commuters, the path choice parameter will be modified as $\boldsymbol{\theta}(\boldsymbol{x}, \boldsymbol{N}, \lambda)$, indicating that commuters' path choices depend on transit frequencies $\boldsymbol{x}$, AMoD fleet size $\boldsymbol{N}$ and the discount factor $\lambda$.
And the optimization with design-dependent choices can be formulated as
\begin{equation}
\label{eq:complete_problem}
    \min_{\boldsymbol{x} \in \Bar{\mathcal{X}}, \boldsymbol{N} \in \Bar{\mathcal{N}}, \lambda \in [\underline{\lambda}, \Bar{\lambda}]} \Tilde{Q}(\boldsymbol{x}, \boldsymbol{N}, \boldsymbol{\theta}(\boldsymbol{x}, \boldsymbol{N}, \lambda)).
\end{equation}

The probability of selecting a specific route, denoted by $\boldsymbol{\theta}(\boldsymbol{x}, \boldsymbol{N}, \lambda)$, is influenced by the utility that a commuter derives from each available route option. This utility, represented by $\mu_t^{u,v,r}(\boldsymbol{x}, \boldsymbol{N}, \lambda)$, pertains to a commute $(u, v)$, departing at time $t$ and choosing route $r$, within the context of transit schedules $\boldsymbol{x}$, AMoD fleet allocations $\boldsymbol{N}$, and discount rate $\lambda$ for AMoD services. It is postulated that the utilities associated with different route options are predominantly affected by two factors: i) \emph{journey time}, and ii) \emph{monetary cost}.

For a route $r \in \mathcal{R}^{u,v}$, the journey time consists of in-vehicle travel time, waiting time, and walking time. The in-vehicle travel time can be formulated as $\sum_{j \in \mathcal{J}(r)} \tau_j$.
For a transit line $l$, the waiting time can be denoted as $\Delta_T / 2x_t^l$ assuming a uniformly distributed headway. The walking time $\tau_{r}^{walk}$ is predetermined for any route $r$ assuming a constant walking speed $\Bar{v}_{walking}$. 
For AMoD services, the waiting time for AMoD services around station $s$ is $\frac{\alpha_s}{\Bar{v}} \sqrt{A_s / N_t^s}$~\citep{Urban_OR}.

% For commuters' comfort level, a convex and differentiable function proposed by \citet{Bertsimas_Sian_Yan_2020} is utilized. The systematic utility for commute $(u,v)$ with route $r$ at time $t$ under decisions $(\boldsymbol{x}, \boldsymbol{N}, \boldsymbol{p})$ can then be represented and a discrete choice model can be used to formulate the design-dependent choices $\boldsymbol{\theta}(\boldsymbol{x}, \boldsymbol{N}, \boldsymbol{p})$ based on utilities, which involves nonlinear formulations.

% We utilized a convex and differentiable function proposed by \citet{Bertsimas_Sian_Yan_2020} to account for the discomfort level in transit lines:
% \begin{equation}
%     \psi(\kappa) =\begin{cases}
%         \kappa & \quad \text{if } \kappa \leq 1,\\
%         e^{\kappa - 1} & \quad \text{if } \kappa > 1,
%     \end{cases}
% \end{equation}
% where $\psi(\kappa)$ stands for a discomfort function and $\kappa$ indicates a ratio between passenger loading and \emph{soft} vehicle capacity.
% The soft vehicle capacity for a transit line $l \in \mathcal{L}$ is $\Bar{K}^l \leq K^l$. 
% For a transit leg $j$ at time $t$, the discomfort level can be represented as $\psi(\frac{O_{l(j),s(j),t}(\boldsymbol{z})}{\Bar{K}^{l(j)} x_t^{l(j)}})$.
% For the model simplicity, we assume that the discomfort level of each transit leg is represented by the discomfort level of the first transit line segment in the leg.
% For AMoD services, let $\pi$ indicate the discomfort level where $\pi < 0$ implies the reduction of discomfort when switching from an empty transit line to an AMoD vehicle.  

For the monetary cost of route $r \in \mathcal{R}^{u,v}$, it is previously defined by $p^{u,v,r}(\lambda)$. Assuming that the commuter utility function is linear in time and cost attributes, the formulation of $\mu_t^{u,v,r}(\boldsymbol{x}, \boldsymbol{N}, \lambda)$ is given by 
\begin{align}
\small
\label{eq:utility}
    & \mu_t^{u,v,r}(\boldsymbol{x}, \boldsymbol{N}, \lambda) = - \beta_2 \cdot p^{u,v,r} (\lambda) \notag \\
    & -\beta_1 \left[\sum_{j \in \mathcal{J}^{transit}(r)} \left( \frac{\Delta_T}{2x_t^{l(j)}} + \tau_j \right) + \sum_{j \in \mathcal{J}^{AMoD}(r)} \left( \frac{\alpha_{s(j)}}{\Bar{v}}\sqrt{\frac{A_{s(j)}}{N_t^{s(j)}}} + \tau_j \right) + \tau_{r}^{walk} \right], %  + \sum_{j \in \mathcal{J}^{global}(r)} \left( \omega_{global} + \tau_j \right) , %- \beta_3 \left[ \sum_{j \in \mathcal{J}^{PT}(r)} \frac{d_j}{\sum_{j \in \mathcal{J}(r)} d_j} \cdot \psi \left( \frac{O_{l(j),s(j),t}(\boldsymbol{z})}{\Bar{K}^{l(j)} x_t^{l(j)}} \right) + \sum_{j \in \mathcal{J}^{L}(r) \cup \mathcal{J}^{G}(r)} \frac{d_j \pi}{\sum_{j \in \mathcal{J}(r)} d_j} \right],
\end{align}
where $\beta_1$ stands for the marginal utility of time, $\beta_2$ represents the marginal utility of money, $l(j)$ denotes the transit line corresponding to transit leg $j$, and $s(j)$ corresponds to the station near the origin of the AMoD leg $j$.  

\subsubsection{Discrete choice models}

With the systematic utility function for different route choices, we can establish a discrete choice model to calculate the design-dependent choices $\boldsymbol{\theta}(\boldsymbol{x}, \boldsymbol{N}, \lambda)$ based on utilities $\mu_t^{u,v,r}(\boldsymbol{x}, \boldsymbol{N}, \lambda)$. For the standard multinomial logit model~\cite{MCFADDEN1974303, Ben-Akiva2003}, a Gumbel-distributed random noise component $\varepsilon^{u,v,r}_t$ is attached to the utility function, i.e.,
\begin{equation}
    \Tilde{\mu}_t^{u,v,r}(\boldsymbol{x}, \boldsymbol{N}, \lambda) = \mu_t^{u,v,r}(\boldsymbol{x}, \boldsymbol{N}, \lambda) + \varepsilon^{u,v,r}_t,
\end{equation}
and the probability for commuters to choose routes with the choice probabilities are given by
\begin{equation} \label{eq:mnl_model}
    \boldsymbol{\theta}(\boldsymbol{x}, \boldsymbol{N}, \lambda) = \frac{\exp{\left( \mu_t^{u,v,r}(\boldsymbol{x}, \boldsymbol{N}, \lambda)\right)}}{\sum_{r' \in \mathcal{R}^{u,v}} \exp{\left( \mu_t^{u,v,r'}(\boldsymbol{x}, \boldsymbol{N}, \lambda) \right)}}.
\end{equation}

\begin{figure}[!h]
\centering
\includegraphics[scale=0.45]{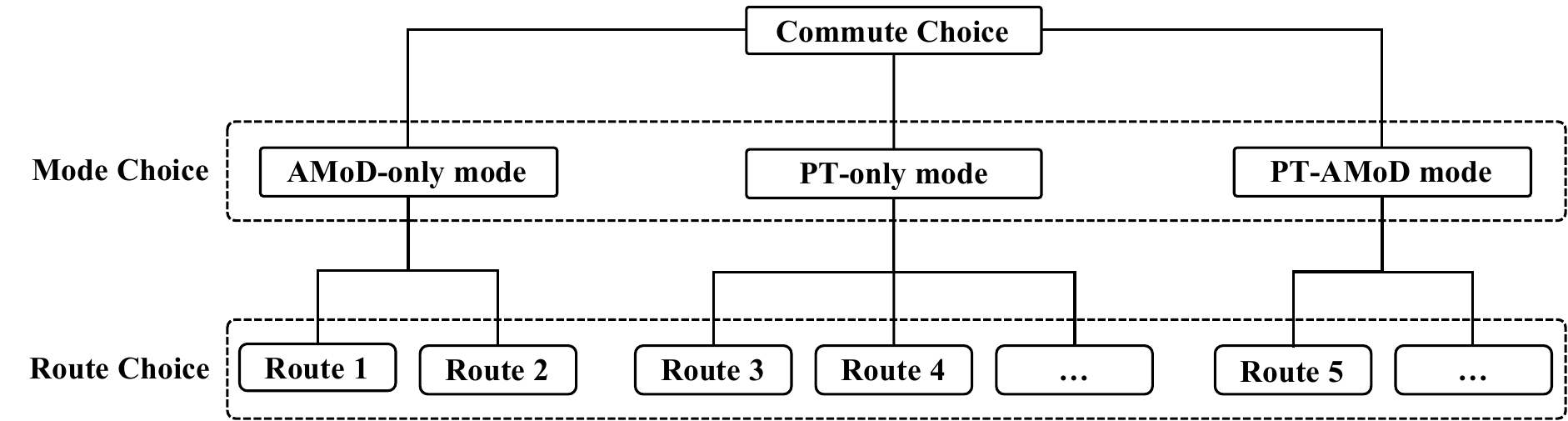}
\caption{Multidimensional choices of the nested logit model.}
\label{fig:NLM}
\end{figure}

The multinomial logit model $(\ref{eq:mnl_model})$ suffers from a property known as the independence from irrelevant alternatives (IIA) as utilities of different routes with identical transportation modes could share similar attributes. To address this issue, we propose a nested logit model, which consists of a two-level choice model: mode choice and route choice. Figure \ref{fig:NLM} displays the two-level decisions that commuters has to make in the nested logit model. Commuters first choose the mode between AMoD-only mode, PT-only mode, and PT-AMoD mode. Under each mode, commuters then make a route choice given a set of available routes with the selective mode.

For a set of route options $\mathcal{R}^{u,v}$ for a commute $(u,v)$, let $\mathcal{R}^{u,v}_{P}$, $\mathcal{R}^{u,v}_{A}$ and $\mathcal{R}^{u,v}_{PA}$ represent the set of route options corresponds to PT-only mode, AMoD-only mode, and PT-AMoD mode, respectively. Then the route choice probability can be formulated as
\begin{equation}
\label{eq:theta_NLM}
\small
    \theta_t^{u,v,r}(\boldsymbol{x}, \boldsymbol{N}, \lambda) = \left[ \frac{\exp(\phi_{m}\mu_t^{u,v,r}(\boldsymbol{x}, \boldsymbol{N}, \lambda))}{\sum_{r \in \mathcal{R}_m^{u,v}}\exp(\phi_{m}\mu_t^{u,v,r}(\boldsymbol{x}, \boldsymbol{N}, \lambda))} \right] \cdot \left[  \frac{\exp(\phi I_m)}{\exp(\phi I_P) + \exp(\phi I_A) + \exp(\phi I_{PA})} \right],
\end{equation}
where $m$ corresponds to the travel mode of route $r$, i.e., $r \in \mathcal{R}^{u,v}_m$, and
\begin{equation}
\small
    I_{m'} = \frac{1}{\phi_{m'}}\ln\left(\sum_{r \in \mathcal{R}_{m'}^{u,v}} \phi_{m'} \mu_t^{u,v,r}(\boldsymbol{x}, \boldsymbol{N}, \lambda)\right), \quad \forall {m'} \in \{\text{P}, \text{A}, \text{PA}\},
\end{equation}
where $\phi$, $\phi_P$, $\phi_A$ and $\phi_{PA}$ are parameters for the nested logit model that have to be estimated from commuter survey data in practice. 

% where parameters $\alpha$, $\alpha_P$, $\alpha_A$ and $\alpha_{PA}$ reflect the correlation of the total utilities for any pair of routes with the same mode. 
% For instance, the correlation for any pair of routes with PT-only mode becomes larger when the value of $\frac{\alpha}{\alpha_P}$ gets smaller.
% In practice, values of parameters in the nested logit model can be estimated through survey data.

\subsubsection{Solution algorithm}

The problem (\ref{eq:complete_problem}) which solves the system design of the TCMUM-AMoD with design-dependent path choices is a nonlinear optimization problem given the nonlinear objective function (\ref{eq:base_problem_shared}a) and the nonlinear discrete choice model formulations (\ref{eq:mnl_model}) and (\ref{eq:theta_NLM}). 
To solve this problem, we utilize a first-order approximation method proposed by \citet{Bertsimas_Sian_Yan_2020}, where the nonlinear function is replaced with a series of locally linear approximations. 
Given a feasible system design point $(\Tilde{\boldsymbol{x}}, \Tilde{\boldsymbol{N}}, \Tilde{\lambda})$, the nonlinear route choice probabilities can be approximated as
\begin{equation}
\label{eq:approx_theta}
\small
    \hat{\theta}_t^{u,v,r}(\boldsymbol{x}, \boldsymbol{N}, \lambda; \Tilde{\boldsymbol{x}}, \Tilde{\boldsymbol{N}}, \Tilde{\lambda}) \approx \theta_t^{u,v,r}(\Tilde{\boldsymbol{x}}, \Tilde{\boldsymbol{N}}, \Tilde{\lambda}) + \bigtriangledown \theta_t^{u,v,r}(\Tilde{\boldsymbol{x}}, \Tilde{\boldsymbol{N}}, \Tilde{\lambda})' (\boldsymbol{x} - \Tilde{\boldsymbol{x}}, \boldsymbol{N} - \Tilde{\boldsymbol{N}}, \lambda - \Tilde{\lambda}),
\end{equation}
where the new system design point $(\boldsymbol{x}, \boldsymbol{N}, \lambda)$ is close to the initial point $(\Tilde{\boldsymbol{x}}, \Tilde{\boldsymbol{N}}, \Tilde{\lambda})$. By substituting the approximated path choice probability (\ref{eq:approx_theta}) into the optimization (\ref{eq:complete_problem}), we obtain a linear optimization problem. Given a feasible system design point $(\boldsymbol{x}^{(i-1)}, \boldsymbol{N}^{(i-1)}, \lambda^{(i-1)})$, a locally better system design can be solved with the problem
\begin{subequations}
\label{eq:iterative_problem}
\begin{align}
\small
    \min_{\boldsymbol{x}, \boldsymbol{N}, \lambda} \quad & \Tilde{Q} \left( \boldsymbol{x}, \boldsymbol{N}, \hat{\boldsymbol{\theta}}(\boldsymbol{x}, \boldsymbol{N}, \lambda; \boldsymbol{x}^{(i-1)}, \boldsymbol{N}^{(i-1)}, \lambda^{(i-1)}) \right) \\
    \text{s.t.} \quad 
    & \boldsymbol{x}^{(i-1)} - \boldsymbol{\rho} \leq \boldsymbol{x} \leq \boldsymbol{x}^{(i-1)} + \boldsymbol{\rho}, \\
    & \boldsymbol{N}^{(i-1)} - \boldsymbol{\eta} \leq \boldsymbol{N} \leq \boldsymbol{N}^{(i-1)} + \boldsymbol{\eta}, \\
    & \lambda^{(i-1)} - \sigma \leq \lambda \leq \lambda^{(i-1)} + \sigma, \\
    & \boldsymbol{x} \in \Bar{\mathcal{X}},\\
    & \boldsymbol{N} \in \Bar{\mathcal{N}}, \\
    & \underline{\lambda} \leq \lambda \leq \Bar{\lambda},
\end{align}
\end{subequations}
where constant vectors $\boldsymbol{\rho}$, $\boldsymbol{\eta}$ and $\sigma$ specify step sizes for line frequencies $\boldsymbol{x}$, local AMoD fleet allocations $\boldsymbol{N}$ and discount rate $\lambda$, respectively. For the nonlinear objective function (\ref{eq:base_problem_shared}a), given we are solving the problem iteratively, we approximate the objective function of iteration $i$ by using system design decisions $(\boldsymbol{x}^{(i-1)}, \boldsymbol{N}^{(i-1)}, \lambda^{(i-1)})$ in iteration $i-1$. Therefore, the problem (\ref{eq:iterative_problem}) becomes a linear program that can be solved efficiently. 

The heuristic algorithm for solving the network design of the TCMUM-AMoD system with design-dependent path choices is described in Algorithm \ref{alg:1}.
\begin{algorithm}[!h]
\caption{First-order approximation algorithm for solving the system design of the TCMUM-AMoD with design-dependent path choices. 
\textbf{Input}: initial feasible system design point $(\boldsymbol{x}^{(0)}, \boldsymbol{N}^{(0)}, \lambda^{(0)})$, step size vectors $\boldsymbol{\rho}$, $\boldsymbol{\eta}$ and $\sigma$, termination threshold $\epsilon$, maximum iteration $\kappa$. 
\textbf{Output}: locally optimal system design $(\boldsymbol{x}^{*}, \boldsymbol{N}^{*}, \lambda^{*})$. }
\label{alg:1}
\begin{algorithmic}[1]
\Function{First-Order-Approximation}{$(\boldsymbol{x}^{(0)}, \boldsymbol{N}^{(0)}, \lambda^{(0)})$, $\boldsymbol{\rho}$, $\boldsymbol{\eta}$, $\sigma$, $\epsilon$}
    \State $i \gets 1$
    \State $\Tilde{Q}_{prev} \gets 0$
    \While {$i \leq \kappa$}
        \State Solve problem (\ref{eq:iterative_problem}) with a feasible point $(\boldsymbol{x}^{(i-1)}, \boldsymbol{N}^{(i-1)}, \lambda^{(i-1)})$, approximated objective functions, step sizes $\boldsymbol{\rho}, \boldsymbol{\eta}, \sigma$ and get the optimal solution $(\boldsymbol{x}^{(i)}, \boldsymbol{N}^{(i)}, \lambda^{(i)})$ and the objective value $\Tilde{Q}^*$
        \State $\textit{threshold} \gets |\Tilde{Q}^* - \Tilde{Q}_{prev}|$
        \If {$\textit{threshold} \leq \epsilon$}
            \State \textbf{break}
        \Else
            \State $i \gets i + 1$
            \State $\Tilde{Q}_{prev} \gets \Tilde{Q}^*$
        \EndIf
    \EndWhile
    \State \textbf{return} $(\boldsymbol{x}^{(i)}, \boldsymbol{N}^{(i)}, \lambda^{(i)})$
\EndFunction
\end{algorithmic}
\end{algorithm}

It is worth noting that the Algorithm \ref{alg:1} is guaranteed to output a locally optimal solution instead of the global optimum. To improve the quality of solutions, we generate multiple starting points $(\boldsymbol{x}^{(0)}, \boldsymbol{N}^{(0)},\lambda^{(0)})$, run the approximated algorithm multiple times and select the best system design solution.
Regarding the step size vectors $\boldsymbol{\rho}$, $\boldsymbol{\eta}$ and $\sigma$, they have to be chosen to balance the computation complexity and algorithm accuracy. A smaller step size leads to a more accurate local optimal solution, but it will take more iterations for the algorithm to converge. 
For the gradient $\bigtriangledown \theta_t^{u,v,r}(\Tilde{\boldsymbol{x}}, \Tilde{\boldsymbol{N}}, \Tilde{\lambda})$ in the equation (\ref{eq:approx_theta}), it can be computed using the automatic differentiation approach.

% \subsubsection{System optimum design}

% Finally, we introduce an optimal system design for the TCMUM-AMoD framework. Within this model, we permit the route selection probabilities, $\boldsymbol{\theta}$, to encompass any legitimate probability distribution across the route alternatives. Here, commuters are guided to select routes that enhance the overall system efficiency. In practical terms, achieving such a system optimum can be facilitated through incentives that encourage commuters to opt for the route that provides the greatest benefit to the system as a whole. The system optimum system design can be generated by

% \begin{subequations}\label{eq:system_optimum}
% \begin{align}
%      \min_{\boldsymbol{x} \in \Bar{\mathcal{X}}, \boldsymbol{N} \in \Bar{\mathcal{N}}, \boldsymbol{\theta}} \quad & \Tilde{Q}(\boldsymbol{x}, \boldsymbol{N}, \boldsymbol{\theta}) \\
%      \text{s.t.} \quad \quad \;  & \sum_{r \in \mathcal{R}^{u,v}} \theta_t^{u, v, r} = 1, \quad \forall u,v \in \mathcal{U}, t \in \mathcal{T} \\
%      & \quad \boldsymbol{\theta} \geq 0.
% \end{align}
% \end{subequations}

\section{Numerical Experiments}
\label{sec:results}

In this section, we conduct numerical experiments on the Chicago transit network operated by the Chicago Transit Authority (CTA), which is one of the largest transit system in the North America. All models are implemented in Python programming language~\cite{PYTHON} and solved using Gurobi 10.0.2~\cite{gurobi}. All experimental results were generated on a machine with a 3.0 GHz AMD Threadripper 2970WX Processor and 128 GB Memory.

Before delving into the details of our numerical experiments, it is important to clarify that the aim of these experiments is not to offer policy recommendations to transit authorities. Instead, the objective is to demonstrate the applicability of our proposed methodology using realistic data. Given the constraints of the data available to us, we employed the multinomial logit model (\ref{eq:mnl_model}) to simulate the route choice behavior of commuters. It should be noted that while the parameters used in this study may not be precise, they are considered sufficiently reasonable to provide valuable insights.

\subsection{Data description}

\begin{table*}[p!]
    \caption{Model parameters and values.}
    \label{tab:model_parameter}
    \centering
    \begin{adjustbox}{max width=0.9\textwidth}
    \begin{tabular}{ l | l | c }
     \hline 
     Parameter & Explanation & Value\\
     \hline \hline 
     \multicolumn{3}{c}{\textbf{Network Design Model Parameters}}\\
     \hline
     $T_{start}$ & Start time of planning period & 06:00 \\
     $T_{end}$ & End time of planning period & 10:00 \\
     $T$ & Number of time periods & 48 \\
     $\Delta_T$ & Length of each time interval & 5 (min) \\
     $|\mathcal{L}^B|$ & Number of bus lines & 40 \\
     $|\mathcal{L}^R|$ & Number of rail lines & 1 \\
     $|\mathcal{U}|$ & Number of morning commutes & 2276 \\ 
     $\underline{B}_R$ & Minimum number of departures for rail & 0.5\\
     $\Bar{B}_R$ & Maximum number of departures for rail & 2.5\\
     $\Bar{B}_B$ & Maximum number of departures for bus & 1\\
     $B_{bus}$ & Number of available bus vehicles & 814\\
     $B_{rail}$ & Number of available rail vehicles& 94 \\
     $K^{l}, l \in \mathcal{L}^B$ & Bus vehicle capacity & 70 \\
     $K^{l}, l \in \mathcal{L}^R$ & Rail vehicle capacity & 640 \\
     $|\mathcal{S}|$ & Number of rail stations & 1 \\
     $A$ & Area of the nearby region for rail station & 90 $(km^2)$\\
     $\alpha$ & Coefficient for approximating local trip distance & 0.667 \\
     $\Bar{v}$ & Average vehicle speed & 20 (mph) \\
     $\delta_w$ & Maximum wait time for FMLM sharing AMoD trips & 60 (seconds) \\
     $\delta_d$ & Maximum delay time for FMLM sharing AMoD trips & 60 (seconds) \\
     \hline \hline
     \multicolumn{3}{c}{\textbf{Discrete Choice Model Parameters}}\\
     \hline 
     $\underline{\lambda}$ & Minimum discount rate for AMoD services & 0.1 \\
     $\Bar{\lambda}$ & Maximum discount rate for AMoD services & 1 \\
     $f^l$ & Fare for transit system & 2.5 (dollars)\\
     $\nu$ & Discount factor for transfers in transit & 0 \\
     $f^{base}$ & Base fare for AMoD services & 1.87 (dollars)\\
     $f^{book}$ & Booking fare for AMoD services & 1.85 (dollars)\\
     $f^{min}$ & Minimum fare for AMoD services & 4.98 (dollars)\\
     $\pi_d$ & Distance fare rate for AMoD services & 0.85  (dollars/mile)\\
     $\pi_t$ & Time fare rate for AMoD services & 0.30 (dollars/minute)\\
     $\Bar{v}_{walking}$ & Average walking speed & 3 (mph) \\
     $\beta_1 (AMoD)$ & Marginal utility of time in AMoD & 16.3 (dollars/hour)\\
     $\beta_1 (transit)$ & Marginal utility of time in transit & 21.1 (dollars/hour)\\
     $\beta_2$ & Marginal utility of money & 1 \\
     \hline \hline
     \multicolumn{3}{c}{\textbf{First-Order Approximation Algorithm Parameters}}\\
     \hline 
     $\epsilon$ & Termination threshold & 0.1 \\
     $\kappa$ & Maximum iteration & 15 \\
     $\rho \; (rail)$ & Step size for rail frequencies & 0.1 \\
     $\eta$ & Step size for AMoD allocations & 10 \\
     $\sigma$ & Step size for AMoD discount rate & 0.1 \\
     \hline
     \end{tabular}
     \end{adjustbox}
\end{table*}

\begin{figure*}[!h]
    \centering
    \includegraphics[width=0.95\textwidth]{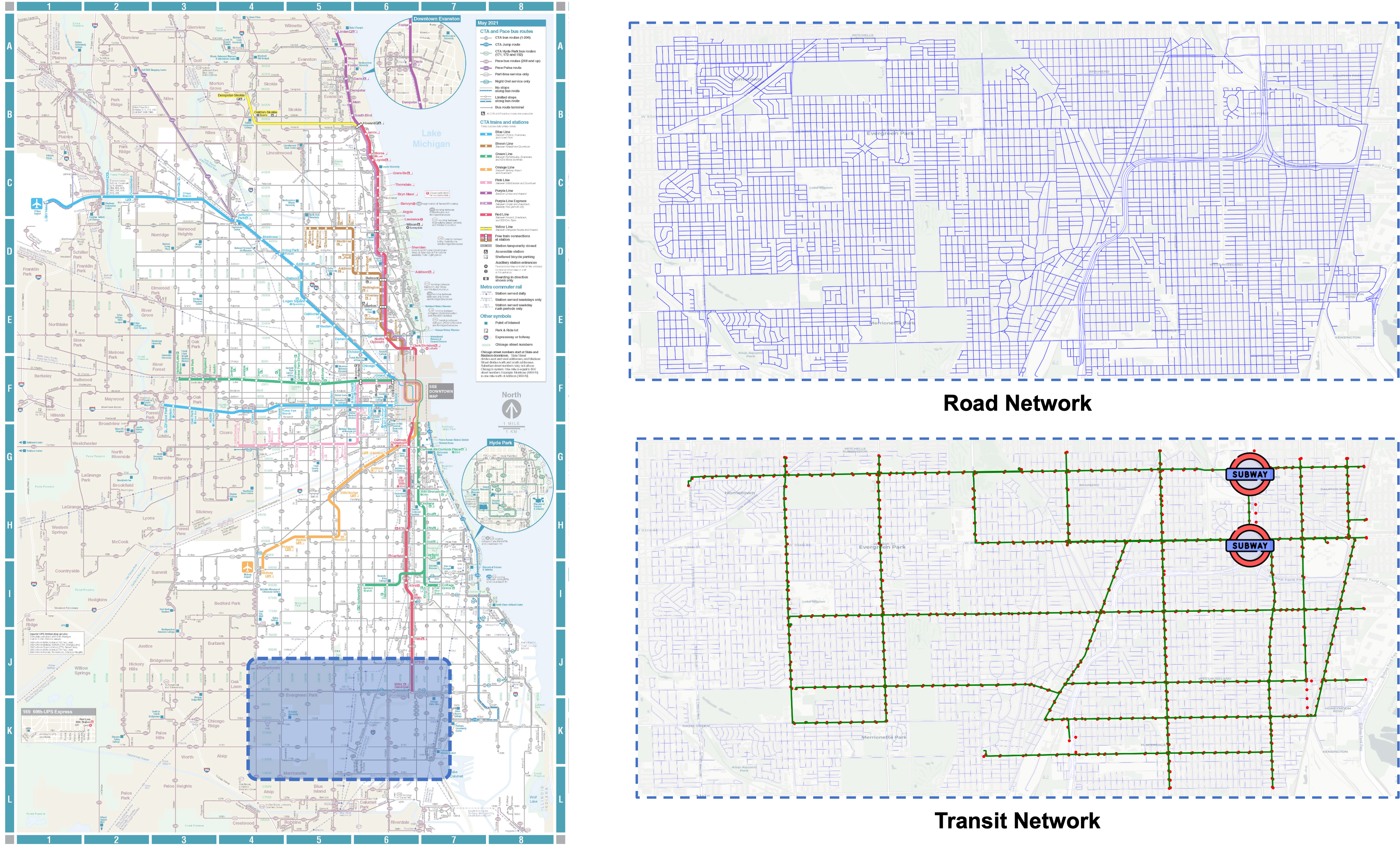}
    \caption{Road and transit networks for the study region. Blue region indicates the study region within the CTA network, blue lines represent road network, green lines denote bus network, red dots stand for bus stops, and rail symbols are rail stations.}
    \label{fig:area_map}
\end{figure*}

\subsubsection{Transit network data}
The parameter values utilized in the experiments are presented in Table \ref{tab:model_parameter}. The geographical focus of the experiments is the southern region of Chicago, characterized by several bus routes that interface and integrate with the Red line of the CTA's rail network. The road and transit networks under consideration are depicted in Figure \ref{fig:area_map}. Within the specified study area, there are 20 distinct bus routes: Routes 3, 4, 8A, 9, X9, 29, 34, 52A, 53A, 87, 95, 100, 103, 106, 108, 111, 111A, 112, 115, and 119. Given that each bus route operates in two directions, our analysis encompasses a total of 40 bus routes, i.e., $|\mathcal{L}^B| = 40$. The rail service included in this study is the Red line of the CTA transit network. Considering the focus on morning commute patterns, only the inbound direction of the Red line is taken into account for this analysis.

The analysis covers the morning hours from 6 AM to 10 AM, segmenting this time frame into 5-minute intervals, resulting in a total of 48 time periods within the model. For each bus route, it is assumed that there is a maximum of one departure per 5-minute interval. Regarding the rail service, the frequency of departures varies, with headways ranging from 2 to 10 minutes. This variation is designed to ensure both a minimum gap between consecutive vehicles and adherence to a baseline level of service.

The experiments leverage data collected over 20 weekdays in September 2019. The information pertaining to the current transit schedules is obtained from an open-source Generalized Transit Feed Specification (GTFS) dataset, which is regularly updated and released by the CTA on a monthly basis. The available number of bus and rail vehicles are calculated based on the transit schedule information. There are $B_{bus}=814$ bus trips and $B_{rail}=94$ rail trips during the 4-hour study period for transit lines in the model.

The travel times between any two stops, accommodating various service patterns, are computed using data from the Automatic Vehicle Location (AVL) dataset for September 2019, also provided by CTA. For transit vehicle capacity, it is posited that each bus is capable of accommodating up to 70 passengers. Similarly, for the rail system, each 8-car rail vehicle is designed to carry a maximum of 640 passengers. 

\subsubsection{Demand data}
Given our study's focus on a specific subregion within the CTA's network, we categorize the passenger demand data into two distinct types based on their travel patterns. The local demand is defined as passengers whose trip origins and destinations are both located within the boundaries of the study area. Conversely, the downtown demand refers to passengers who begin their journeys within the study area and then utilize the Red Line service for traveling to their destinations outside the study area. 

The demand data employed in the analysis is sourced from CTA's ODX dataset for September 2019. The ``ODX'' stands for the "origin, destination, and transfer inference algorithm," an advanced algorithm developed by Gabriel et al.~\cite{ODX_Gabriel} and currently implemented by the CTA. The CTA utilizes a fare collection system that requires passengers to "tap-on" but does not record their exit points, thereby not capturing alighting data. To bridge this gap, the ODX algorithm is employed to infer passengers' alighting points. Detailed insights into the creation and implementation of the ODX algorithm can be found in \citet{ODX_Gabriel, ODX_Zhao, ROVE_2022}.

In this paper, we assume there exists a total commuter demand of 12,400 individuals throughout the studied period. This analysis incorporates the demand data for both local and downtown commuters as sourced from the CTA, to establish demand seeding matrices. These matrices serve as the foundation for generating the demand data used in the optimization model. In the process of constructing these demand seeding matrices, we calculate the demand for the rail system based on an average over 20 workdays within the month of September 2019. Conversely, the bus system's demand is derived from the data of a specific workday, namely September 5th, chosen due to the bus system's sparse demand patterns. Altogether, the dataset encompasses $2276$ distinct commutes within the analysis, denoted as $|\mathcal{U}| = 2276$. 

\subsubsection{AMoD parameters}
In the context of first-mile AMoD services, our study area includes two rail stations. Due to their proximity, these stations are consolidated into a single entity within our analysis. We determine the optimal fleet size for AMoD in the study region across each time interval. The study region is defined as a rectangular area, approximately $A = 90 \; km^2$ in size. Following the methodology outlined in \citet{Urban_OR}, we adopt a coefficient of $\alpha = 0.667$ for estimating the distance of local trips. Furthermore, the model assumes an average vehicle speed of $\Bar{v} = 20 \; mph$. 

To reduce the computational complexity, we only allow sharing within FMLM AMoD trips. To generate sharing scenarios for the first-mile AMoD services, we assume that each vehicle is limited to being shared by two separate commuters. Additionally, it is crucial that each commuter within a shared route adheres to specified constraints regarding the maximum wait time ($\delta_w$) and the maximum delay time ($\delta_d$). For the purposes of this study, both the maximum wait time and the maximum delay time are established at 60 seconds, i.e., $\delta_w = \delta_d = 60$.

\subsubsection{Discrete choice model parameters}

For the transit system pricing, a uniform fare structure is implemented, with a flat fare of $f^l = 2.5$ dollars for using any transit line $l \in \mathcal{L}$. Additionally, transfers between transit lines are offered at no additional cost, i.e., $\nu = 0$.

In the context of the AMoD services pricing mechanism~\cite{Uber_cost}, the pricing model includes a base fare of $f^{base} = 1.87$ dollars, a booking fee of $f^{book} = 1.85$ dollars, and a minimum fare of $f^{min} = 4.98$ dollars. The fare structure also incorporates a distance-based rate of $\pi_d = 0.85$ dollars per mile and a time-based rate of $\pi_t = 0.30$ dollars per minute. For commuters who take AMoD services, a discount rate $\lambda$ is applied, varying within a range from $\underline{\lambda} = 0.1$ to $\Bar{\lambda} = 1$.

Regarding the parameters in the utility function as specified in Equation $(\ref{eq:utility})$, with $\beta_2$ set to 1, $\beta_1$ represents the value of time in vehicles. For this analysis, the value of time spent in transit vehicles is set at 21.1 dollars per hour, whereas the value of time in AMoD vehicles is determined to be 16.3 dollars per hour, based on findings from \citet{Hyland2018}. The average walking time for commuters is set to be $\Bar{v}_{walking} = 3$ mph.

\subsubsection{Algorithm parameters}

In the application of the first-order approximation algorithm to solve the TCMUM-AMoD system design problem, we initiate the process with 15 randomly selected starting solutions. The algorithm is set to terminate under one of two conditions: either after $\kappa=15$ iterations or when the objective values of two successive iterations are within a tolerance of $\epsilon=0.1$. 

Given the relatively low demand observed for bus routes, the decision variables related to bus schedules are treated as integers. This approach is adopted because employing continuous variables for these schedules often results in very small, practically negligible values, essentially equivalent to discontinuing the bus route. Consequently, continuous step sizes are reserved exclusively for rail decision variables, with a specified step size of $\rho(rail) = 0.1$. 

For the allocation of AMoD fleet, the step size is set at $\eta = 10$, reflecting adjustments in the number of AMoD vehicles allocated. Similarly, the step size for modifying the AMoD service discount rate is $\sigma = 0.1$.

\subsection{Model results}

In the experiments, we would like to understand the trade-offs between different number of buses and AMoD vehicles under different demand profiles. Therefore, we adjust three parameters in the experiments: i) proportion of available bus runs $\gamma$, ii) number of available AMoD vehicles $\Bar{N}$, and iii) proportion of downtown commuters $\psi$. 

In the baseline scenario, we assume the percentage of available bus runs $\gamma$ and the number of available AMoD vehicles $\Bar{N}$ follows two relationships: i) passenger car equivalence (PCE), and ii) capital cost equivalence (CCE). For instance, if we are removing 20\% of bus runs within $B_{bus}=814$ total bus runs during the four-hour study period, it is equivalent to remove around $\frac{814 \times 20\%}{4} \thickapprox 41$ buses assuming each bus makes one run every hour. Then, the passenger car equivalence leads to $82$ AMoD vehicles as the passenger car unit (PCU) for bus is $2.0$~\cite{PCU}. Regarding the capital cost equivalence, $41$ buses is equivalent to $164$ AMoD vehicles, given the cost for a E-bus around $\$800,000$~\cite{bus_cost} the cost for an AMoD vehicle around $\$200,000$~\cite{AMoD_cost}. 

For the proportion of downtown commuters, we assume the baseline number to be $\psi = 80\%$, reflecting commuters' demand pattern in CTA. The demand profile used in the optimization model is generated as follows: i) generate the downtown demand with a demand level of $12,400 \cdot \psi$, ii) generate the local demand with a demand level of $12,400 \cdot (1-\psi)$, iii) combine local and downtown demand as the final demand profile. 

\begin{figure}[!h]
    \centering
    \begin{subfigure}[b]{0.45\textwidth}
        \centering
        \includegraphics[width=\textwidth]{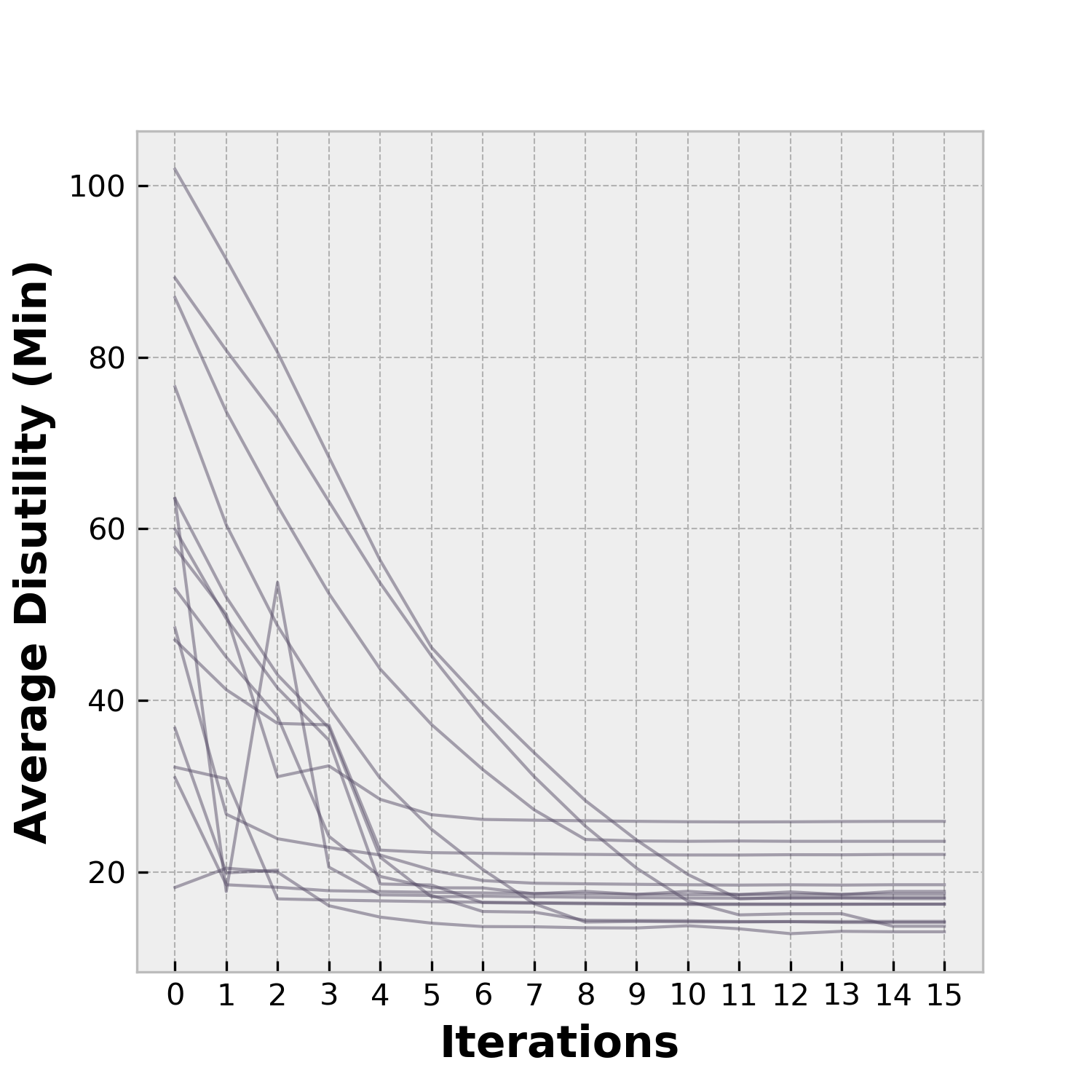}
        \caption[]%
        {Scenarios with 80\% buses, 82 AMoD vehicles, and 80\% downtown commuters.}
        \label{}
    \end{subfigure}
    \begin{subfigure}[b]{0.45\textwidth}  
        \centering 
        \includegraphics[width=\textwidth]{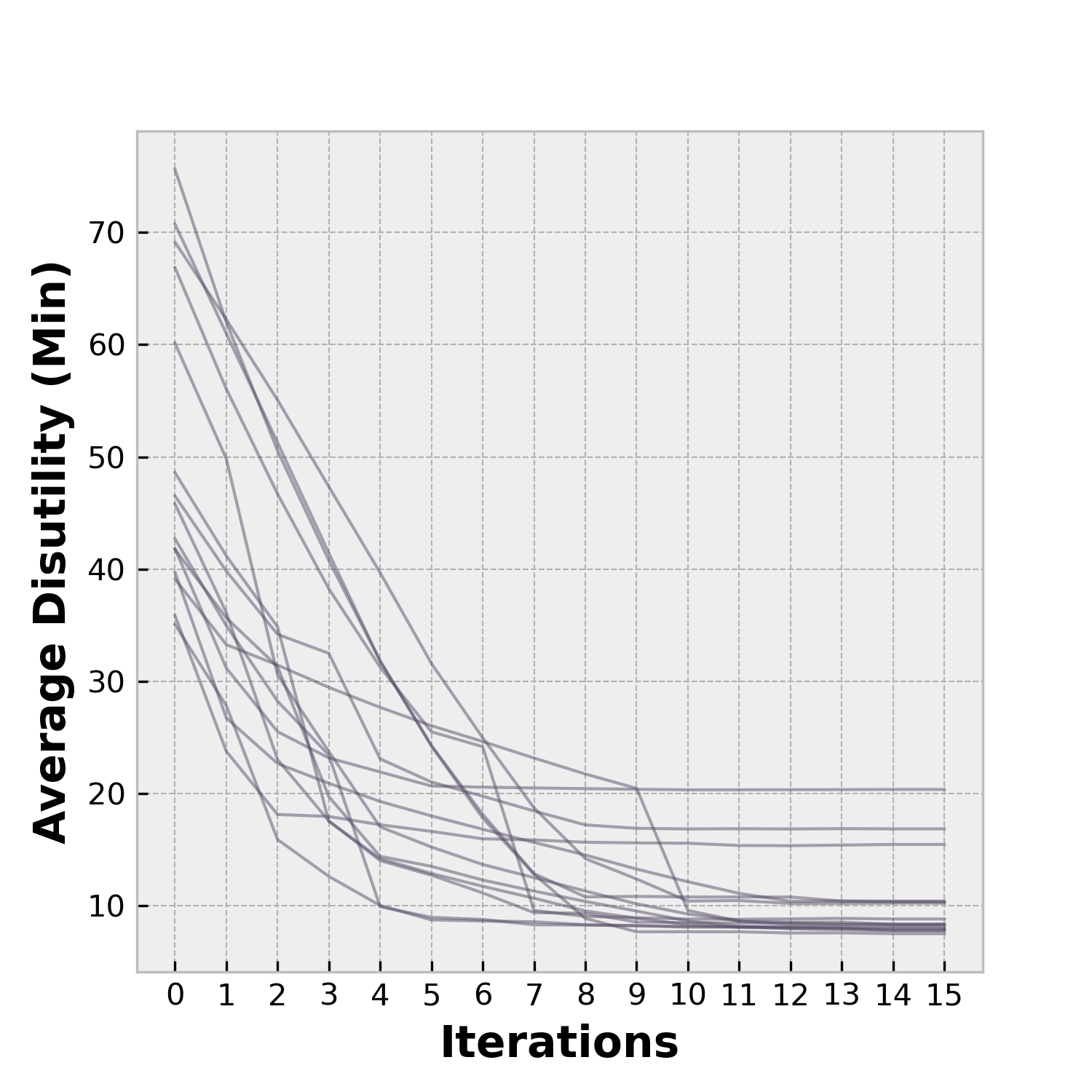}
        \caption[]%
        {Scenarios with 80\% buses, 164 AMoD vehicles, and 80\% downtown commuters.}
        \label{}
    \end{subfigure}
    \caption{Convergence of the first-order approximation algorithm.} 
    \label{fig:convergence_map}
\end{figure}

Figure \ref{fig:convergence_map} shows the convergence performance for the proposed first-order approximation algorithm under two scenarios. Although some randomly-generated initial starting points lead to local optimal solutions, the proposed algorithm with 15 initial starts is capable of generating satisfying system design solutions. Also, 15 iterations is enough for the proposed algorithm to converge.

% convergence map

Table \ref{tab:baseline_results_0.8_PCE} presents the findings from the baseline experiment, focusing on a scenario where 80\% of commuters are heading downtown ($\psi = 0.8$), within the context of passenger car equivalence. The term \emph{Avg disutility} refers to the average disutility for commuters, which is a composite measure including walking time, expected waiting time, and excess waiting time. \emph{Avg walking} provides the average walking duration for each commuter, while \emph{Avg waiting} quantifies the average expected wait time. \emph{Avg utility} reflects the average utility for chosen paths by commuters. The percentage of selected bus routes in the optimally designed system is captured by \emph{Line utilization}, whereas \emph{AMoD utilization} indicates the usage rate of AMoD vehicles throughout the studied period. $\lambda^*$ represents the optimal discount rate applied to AMoD service usage. The table also shows the mode share among downtown and local commuters, with \emph{L} and \emph{DT} signifying local and downtown commuters, respectively. \emph{Unserved\%} highlights the proportion of commuters who were not served by the end of the study period, remaining in wait for the next available vehicle.

\begin{table*}[h!]
    \caption{Experimental results for baseline scenario with 80\% downtown commuters ($\psi = 0.8$) under PCE.}
    \label{tab:baseline_results_0.8_PCE}
    \centering
    \begin{adjustbox}{max width=\textwidth}
    \begin{tabular}{|cc|ccccccc|}
     \hline 
     &&\multicolumn{7}{c|}{\textbf{System Performance Indicators}}\\
     \hline
     $\gamma$ & $\Bar{N}$ & Avg disutility & Avg walking & Avg waiting & Avg utility & Line utilization & AMoD utilization & $\lambda^*$ \\
     \hline
     100\% & 0 & 8.88 & 1.97 & 2.15 & -6.00 & 90\% & - & - \\
     80\% & 82 & 13.02 & 1.88 & 2.13 & -5.95 & 93\% & 14\% & 1.0 \\
     60\% & 164 & 9.46 & 1.32 & 1.85 & -6.41 & 76\% & 94\% & 1.0 \\
     40\% & 246 & 10.06 & 1.17 & 1.78 & -6.53 & 73\% & 95\% &  1.0 \\
     20\% & 328 & 11.32 & 1.03 & 1.60 & -6.60 & 49\% & 100\% &  1.0 \\
     0\% & 410 & 13.08 & 0.82 & 1.35 & -6.74 & - & 100\% & 1.0 \\
     \hline
     \hline
     &&\multicolumn{7}{c|}{\textbf{Mode Share for Downtown and Local Commuters}}\\
     \hline 
     $\gamma$ & $\Bar{N}$ & AMoD\% (L) & Bus\% (L)  & Unserved\% (L) & AMoD+rail (DT) & Bus+rail (DT) & Rail (DT) & Unserved\% (DT) \\
     \hline
     100\% & 0 & 0\% & 100\% & 22.4\% & 0\% & 26\% & 74\% & 0.3\% \\
     80\% & 82 & 11\% & 89\% & 27.9\% & 1\% & 25\% & 74\% & 2.4\% \\
     60\% & 164 & 61\% & 39\% & 0\% & 16\% & 18\% & 66\% & 4.1\% \\
     40\% & 246 & 74\% & 26\% & 0\% & 21\% & 13\% & 66\% & 5.0\% \\
     20\% & 328 & 88\% & 12\% & 0\% & 25\% & 8\% & 67\% & 6.3\% \\
     0\% & 410 & 100\% & 0\% & 0\% & 33\% & 0\% & 67\% & 7.9\% \\
     \hline
     \end{tabular}
     \end{adjustbox}
\end{table*}

Under the passenger car equivalence setting, the replacement of buses with AMoD vehicles does not influence the traffic condition. Therefore, the fixed travel time assumption used in this paper holds. Table \ref{tab:baseline_results_0.8_PCE} reveals the following major findings:

\textbf{(1) Buses are efficient then AMoD vehicles.} In every scenario analyzed, commuters experience the lowest disutility in the setup featuring 100\% buses and no AMoD vehicles. Substituting buses with AMoD vehicles tends to decrease both walking and expected waiting times. However, this adjustment leads to increased delays across the system due to an upsurge in excess waiting time, as a significant number of commuters choose to use AMoD services. While public transit lacks the capacity to offer door-to-door service, its strength lies in its ability to move large groups of people efficiently. As a result, for transporting morning commuters, buses emerge as the more effective mode compared to AMoD vehicles.

\textbf{(2) AMoD vehicles can better serve local demand.} With an increase in the number of AMoD vehicles, there's a noticeable decrease in the percentage of unserved local commuters, whereas the percentage of downtown commuters not served goes up. AMoD vehicles are particularly adept at serving local commuters, owing to the less dense demand patterns in local areas. On the other hand, downtown commuter demand is more destination-focused, making bus transportation exceptionally efficient for serving these concentrated demand patterns.

% results under capital cost equivalence
\begin{table*}[b!]
    \caption{Experimental results for baseline scenario with 80\% downtown commuters ($\psi = 0.8$) under CCE.}
    \label{tab:baseline_results_0.8_CCE}
    \centering
    \begin{adjustbox}{max width=\textwidth}
    \begin{tabular}{|cc|ccccccc|}
     \hline 
     &&\multicolumn{7}{c|}{\textbf{System Performance Indicators}}\\
     \hline
     $\gamma$ & $\Bar{N}$ & Avg disutility & Avg walking & Avg waiting & Avg utility & Line utilization & AMoD utilization & $\lambda^*$ \\
     \hline
     100\% & 0 & 8.88 & 1.97 & 2.15 & -6.00 & 90\% & - & - \\
     80\% & 162 & 7.48 & 1.39 & 1.98 & -6.33 & 66\% & 88\%   & 1.0 \\
     60\% & 328 & 8.30 & 1.31 & 1.94 & -6.41 & 56\% & 61\%   & 1.0 \\
     40\% & 492 & 9.05 & 1.19 & 1.93 & -6.55 & 71\% & 100\%   & 1.0 \\
     20\% & 656 & 11.26 & 1 & 1.56 & -6.64 & 49\% &  66\%  & 1.0 \\
     0\% & 820 & 13.03 & 0.82 & 1.28 & -6.73 & - &  100\%  & 1.0 \\
     \hline
     \hline
     &&\multicolumn{7}{c|}{\textbf{Mode Share for Downtown and Local Commuters}}\\
     \hline 
     $\gamma$ & $\Bar{N}$ & AMoD\% (L) & Bus\% (L)  & Unserved\% (L) & AMoD+rail (DT) & Bus+rail (DT) & Rail (DT) & Unserved\% (DT) \\
     \hline
     100\% & 0 & 0\% & 100\% & 22.4\% & 0\% & 26\% & 74\% & 0.3\% \\
     80\% & 162 & 55\% & 45\% & 0.3\% & 14\% & 21\% & 66\% & 3.0\% \\
     60\% & 328 & 64\% & 36\% & 0.1\% & 16\% & 18\% & 66\% & 3.6\% \\
     40\% & 492 & 75\% & 25\% & 0\% & 19\% & 16\% & 65\% & 4.2\% \\
     20\% & 656 & 89\% & 11\% & 0\% & 26\% & 8\% & 67\% & 6.4\% \\
     0\% & 820 & 100\% & 0\% & 0\% & 33\% & 0\% & 67\% & 7.8\% \\
     \hline
     \end{tabular}
     \end{adjustbox}
\end{table*}

\textbf{(3) Discounting AMoD services might not be a good idea.} Across various scenarios, the optimal discount factor for utilizing AMoD services consistently stands at 1.0, suggesting that no discounts are applied to AMoD services. This decision is intuitive given the limited AMoD vehicles; introducing discounts on AMoD fares would render it the most attractive option for commuters, subsequently causing delays across the entire system. Notably, a noticeable increase in disutility is observed when 20\% of buses are substituted with 82 AMoD vehicles. Lowering AMoD fares under these conditions would only exacerbate the issue, leading to further increases in excess waiting times.

Table \ref{tab:baseline_results_0.8_CCE} presents the results for scenarios featuring 80\% downtown commuters ($\psi=0.8$) under the CCE setting. While the CCE does not preserve the fixed travel time assumption, it offers transit agencies a financially viable pathway for deploying the integrated system. Additional insights gleaned from Table \ref{tab:baseline_results_0.8_CCE} include:

\textbf{(4) Enough AMoD vehicles can effectively compliment the transit network.} While substituting buses with AMoD vehicles in the PCE setting does not yield benefits at the system level, a decrease in average disutility is observed within the CCE scenarios. This suggests that a sufficient number of AMoD vehicles can effectively compliment the transit network. Specifically, replacing 20\% of buses with 162 AMoD vehicles results in a 15.8\% reduction in commuter disutility. Nonetheless, the benefits of such replacements exhibit a non-monotonic pattern; substituting 40\% of buses with 328 AMoD vehicles leads to a lesser improvement, reducing commuter disutility by only 6.5\%. Further replacements diminish system performance, underscoring that while AMoD vehicles can enhance transit networks, maintaining an essential level of transit services is crucial for their capacity to transport large numbers of people efficiently.

\textbf{(5) Utilization of AMoD and bus routes are non-monotonic with respect to AMoD-bus configurations.} Contrary to the expected linear trends in line and AMoD vehicle utilization observed in Table \ref{tab:baseline_results_0.8_PCE}, the data in Table \ref{tab:baseline_results_0.8_CCE} reveal non-linear, or non-monotonic, changes. Intuitively, adding more AMoD vehicles might seem like a straightforward way to replace bus routes that serve local commuters. Yet, as the AMoD fleet size increases, the reduced waiting times for these services become more attractive, drawing an increased number of users towards AMoD. This surge in AMoD demand makes it harder for local commuters to access these services due to the limited AMoD resources. Therefore, the optimal system design maintains a level of bus services for local routes to balance the surge AMoD demand during peak periods.

% transit network map with 100% bus vs. 20% bus
\begin{figure}[!h]
    \centering
    \includegraphics[width=0.7\textwidth]{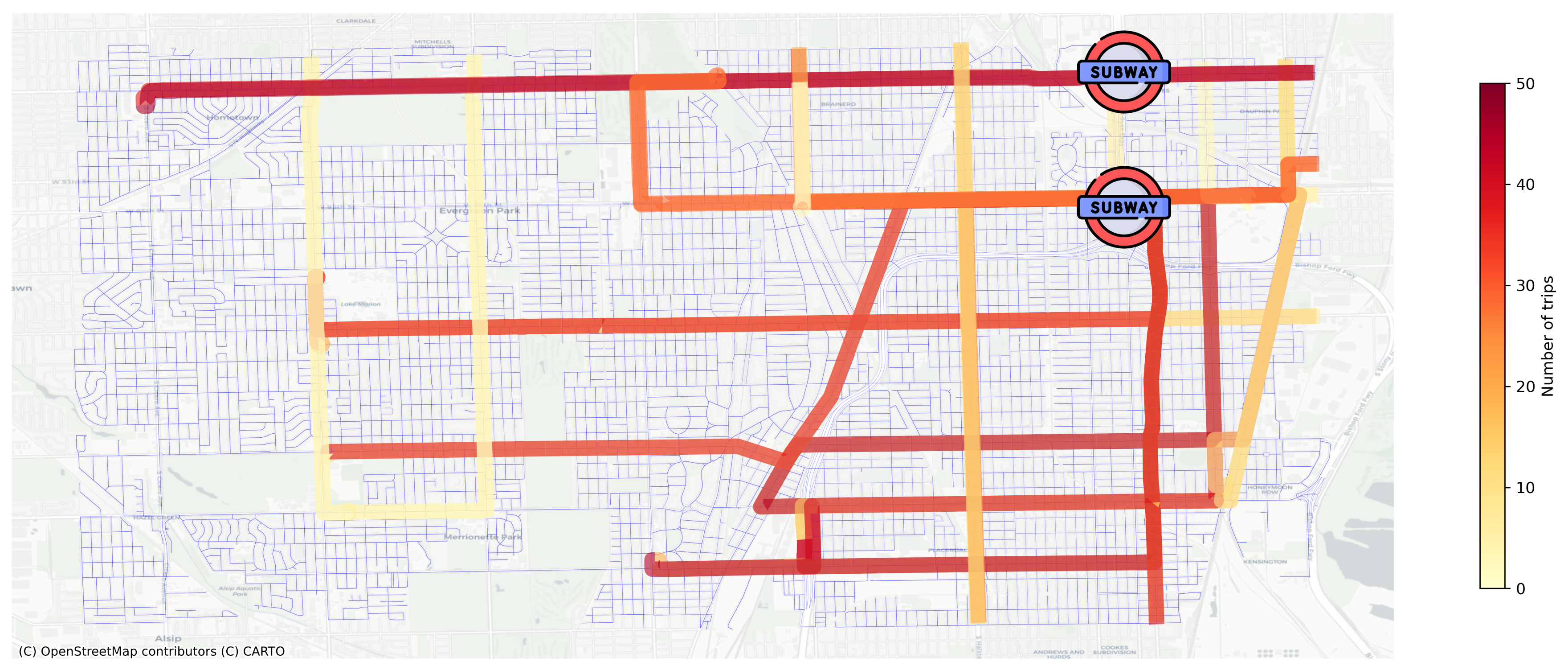}
    \caption{Optimal bus network for 100\% buses, 0 AMoD vehicles, and 80\% downtown commuters.}
    \label{fig:transit_network_100}
\end{figure}

\begin{figure}[!h]
    \centering
    \includegraphics[width=0.7\textwidth]{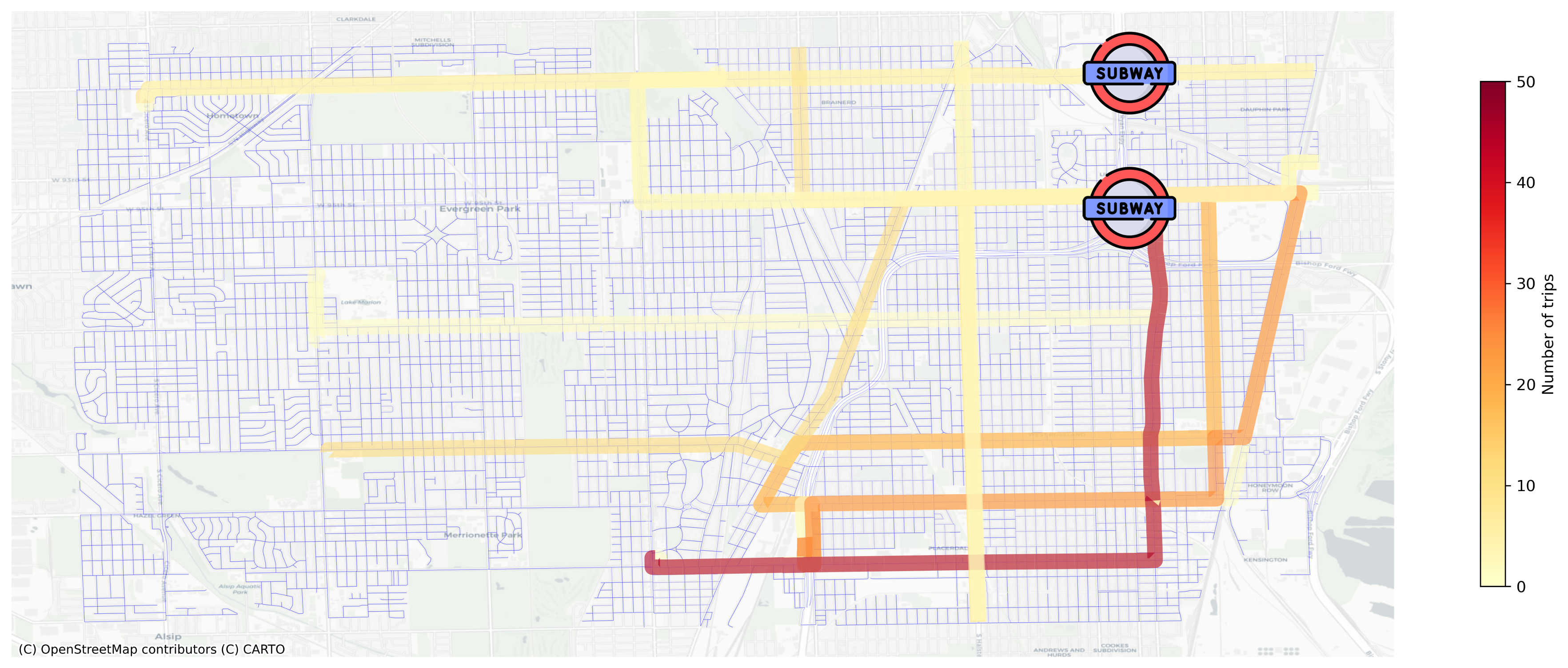}
    \caption{Optimal bus network for 20\% buses, 328 AMoD vehicles, and 80\% downtown commuters under the PCE setting.}
    \label{fig:transit_network_PCE_20}
\end{figure}

\begin{figure}[!h]
    \centering
    \includegraphics[width=0.7\textwidth]{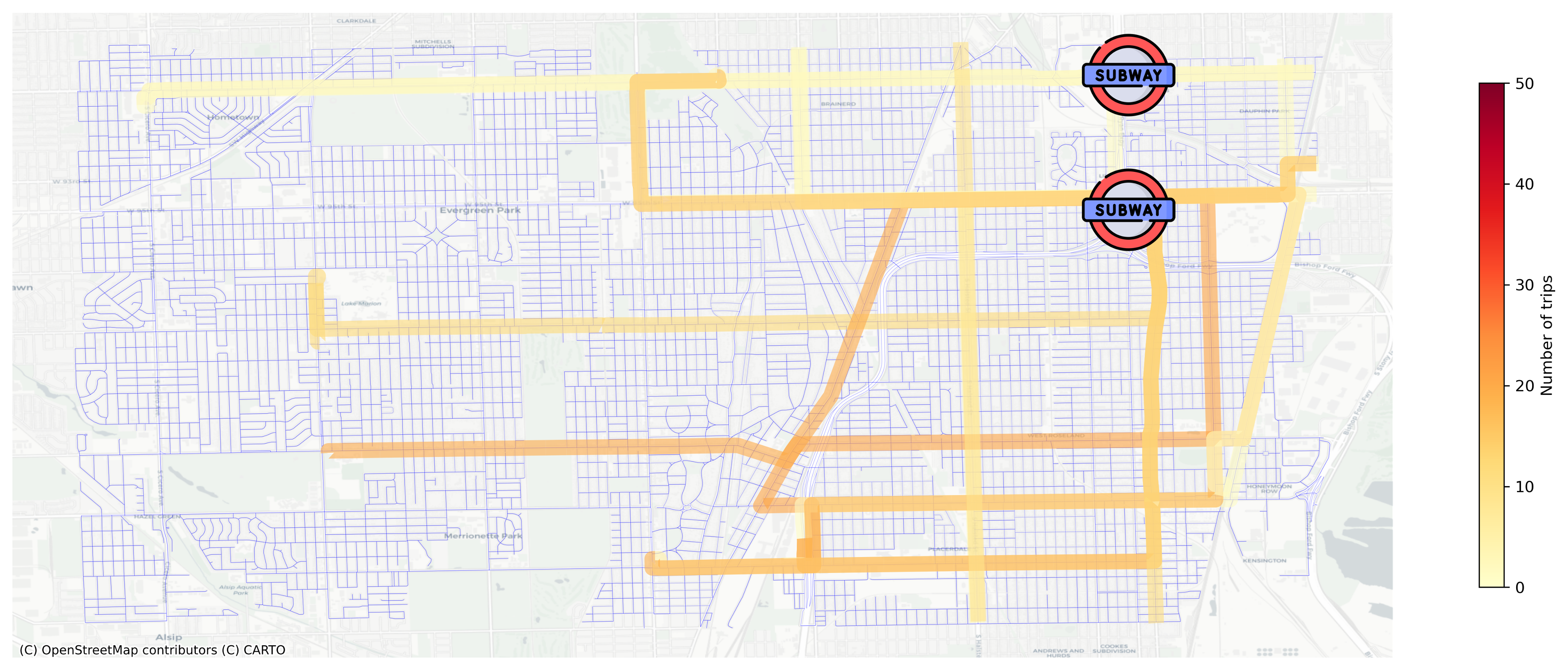}
    \caption{Optimal bus network for 20\% buses, 656 AMoD vehicles, and 80\% downtown commuters under the CCE setting.}
    \label{fig:transit_network_CCE_20}
\end{figure}

Figure \ref{fig:transit_network_100}-\ref{fig:transit_network_CCE_20} illustrate the bus network configuration in a scenario with 80\% downtown commuters. Figure \ref{fig:transit_network_100} details the network design when it's composed entirely of buses (100\%) without any AMoD vehicles. Figure \ref{fig:transit_network_PCE_20} outlines the structure with a significant reduction in buses to 20\%, incorporating 328 AMoD vehicles into the system under the PCE setting. In this transformation, a noticeable shift occurs with the majority of bus routes in the north-south direction being eliminated. However, routes that connect to rail stations are preserved, and these retained bus services enjoy higher frequencies compared to other routes.

Figure \ref{fig:transit_network_CCE_20} shows the bus network configuration in a scenario featuring 20\% buses and 656 AMoD vehicles under the CEE setting. While the same bus routes are preserved as in the PCE scenario, the distribution of buses across these routes is more equitable in the CCE scenario. The increased presence of AMoD vehicles facilitates better service to areas with higher demand, thereby reducing the necessity for bus routes to operate at high frequencies.

\subsection{Sensitivity Analyses}

In this section, we conduct sensitivity analyses over two critical model parameters: i) AMoD fleet size, and ii) percentage of downtown commuters. 

\begin{figure}[!h]
    \centering
    \begin{subfigure}[b]{0.48\textwidth}
        \centering
        \includegraphics[width=\textwidth]{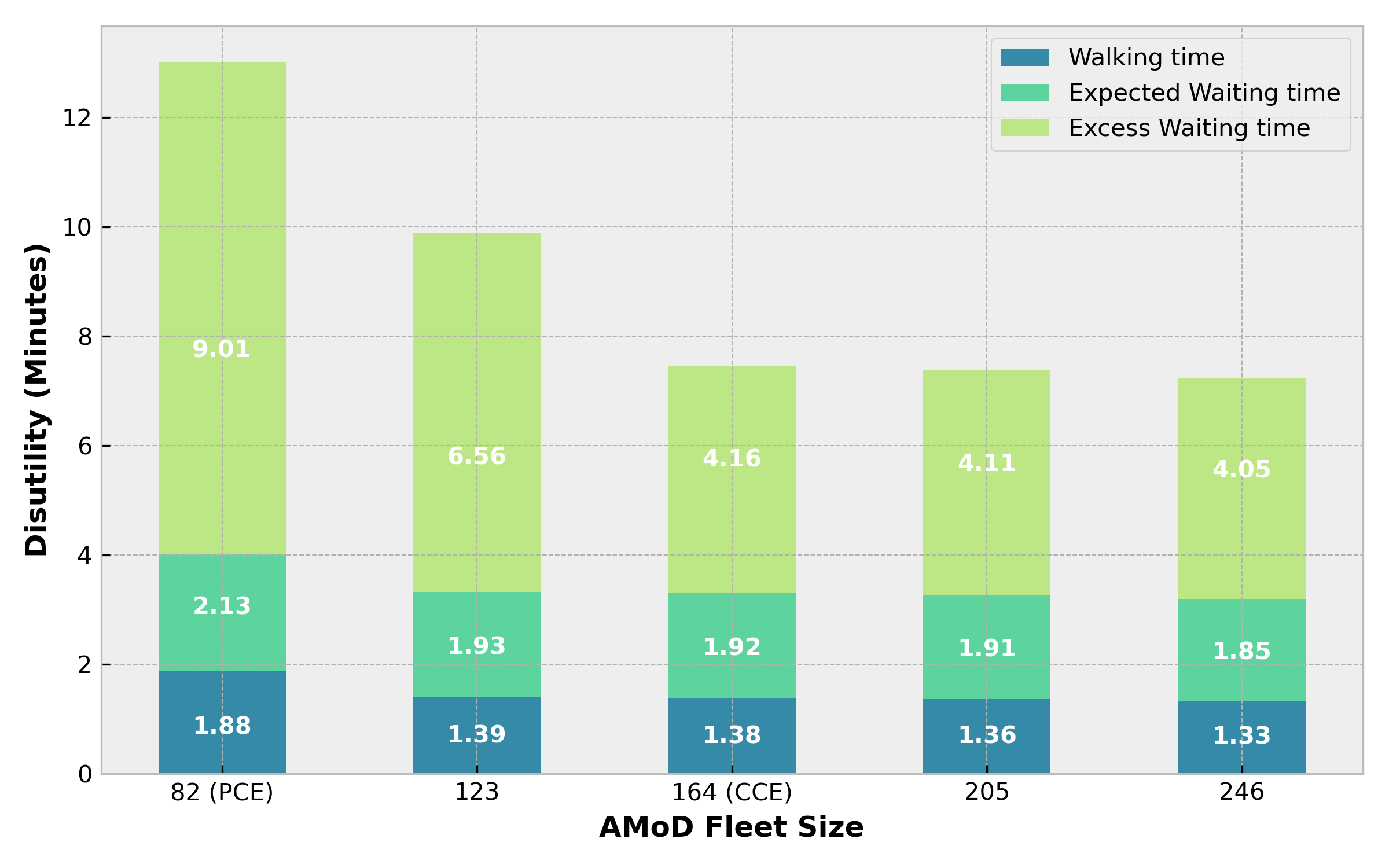}
        \caption[]%
        {Average disutility and its breakdown.}
        \label{fig:fleet_size_a}
    \end{subfigure}
    \begin{subfigure}[b]{0.48\textwidth}  
        \centering 
        \includegraphics[width=\textwidth]{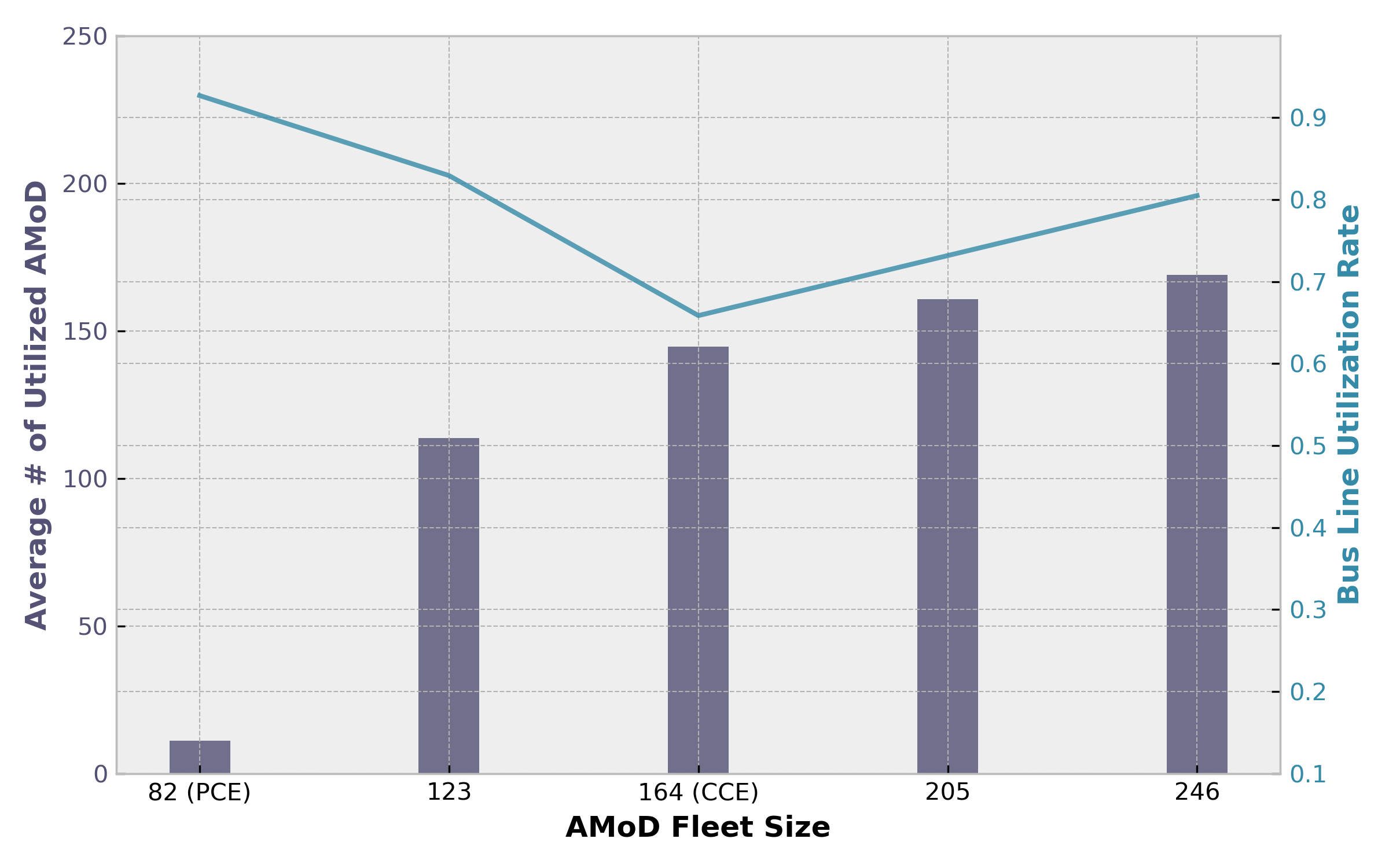}
        \caption[]%
        {Utilization of AMoD vehicles and bus lines.}
        \label{fig:fleet_size_b}
    \end{subfigure}
    \caption{Sensitivity analysis for different AMoD fleet size.} 
    \label{fig:fleet_size}
\end{figure}

Figure \ref{fig:fleet_size} illustrates a sensitivity analysis of varying AMoD fleet sizes, from 82 to 246 vehicles, increasing incrementally by 41, within a scenario that maintains 80\% buses and 80\% downtown commuters. Figure \ref{fig:fleet_size_a} details the average disutility faced by commuters and its breakdown across different AMoD fleet sizes. An increase in the number of AMoD vehicles leads to a decrease in average disutility, attributed to improvements in walking time, expected waiting time, and reduction in excess waiting time. The expansion of the AMoD fleet enhances the provision of door-to-door services, thereby reducing the waiting time for accessing AMoD services.

Figure \ref{fig:fleet_size_b} presents the utilization rates of AMoD vehicles and bus lines, highlighting that the usage of AMoD vehicles rises with an increase in fleet size. Surprisingly, the utilization rate of bus lines initially decreases but then shows an increase. This unexpected trend suggests that while a larger fleet of AMoD vehicles might theoretically replace more bus services—especially those serving local commuters—the decreased waiting time for AMoD services actually encourages more commuters to choose AMoD. Consequently, certain bus routes remain necessary to accommodate commuters who would benefit from AMoD services but cannot access them due to the high demand from others also switching to AMoD.

% 1. fleet size changing

\begin{figure}[!h]
    \centering
    \includegraphics[width=.9\textwidth]{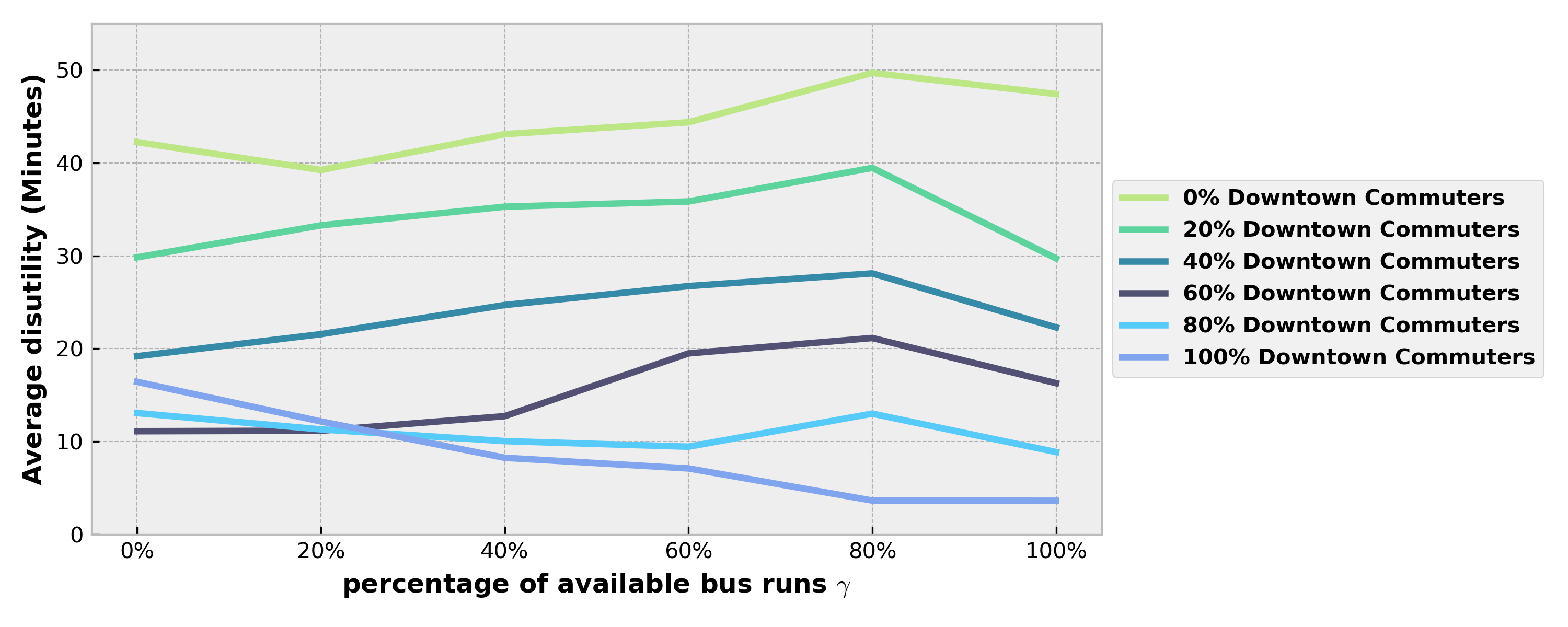}
    \caption{Sensitivity analysis for different percentage of downtown commuters.}
    \label{fig:downtown_ratio}
\end{figure}

Figure \ref{fig:downtown_ratio} shows how the average commuter disutility varies across different ratios of available bus runs ($\gamma$) and the percentage of downtown commuters ($\psi$) under the PCE setting. In scenarios with entirely downtown commuters (100\%), an increase in the proportion of bus services ($\gamma$) correlates with reduced average disutility, as buses are more adept at serving the needs of downtown commuters. Conversely, in scenarios with exclusively local commuters (0\% downtown commuters), a reduction in bus availability generally leads to lower disutility levels. The relationship is not straightforward, however, as a balanced mix of buses and AMoD vehicles can better serve local commuters.

Interestingly, in scenarios without bus services (0\% available bus runs), the situation with 60\% downtown commuters has the lowest level of commuter disutility. In these instances, AMoD vehicles are more efficiently matched to the combined needs of local commuters and downtown commuters traveling to rail stations. Scenarios with 100\% downtown commuters experience surge demand periods that AMoD vehicles alone cannot immediately accommodate, resulting in increased wait times for commuters within the system.

% 2. downtown ratio changing

In conclusion, AMoD vehicles are particularly effective at accommodating local demand patterns, while traditional transit systems excel in serving downtown commuters. Nevertheless, maintaining a synergy between both systems is crucial for addressing the full spectrum of demand patterns. Identifying the sweet spot between the size of the AMoD fleet and the level of transit services emerges as a key strategy for maximizing efficiency across all commuter types.

\section{Conclusions and Future Work}
\label{sec:conclusion}

In this study, we introduce a comprehensive optimization framework designed to jointly optimize the transit networks and frequencies, specify the size and distribution of AMoD fleets, and determine service pricing, all with the objective of minimizing commuters' total disutility. We propose an optimization model for the integrated design of Transit-Centric Multimodal Urban Mobility with Autonomous Mobility-on-Demand (TCMUM-AMoD) systems, which incorporates commuters' mode and route choice behavior via discrete choice models. The proposed optimization model is a mixed integer non-linear program (MINLP) which is intractable to solve at a large scale. Therefore, a first-order approximation algorithm is employed which can solve the problem efficiently. This framework has been tested through a real-world case study in Chicago, encompassing a variety of demand scenarios. The outcomes validate the effectiveness of our model in generating system design solutions. Moreover, the findings reveal the efficiency of AMoD vehicles in meeting local demand patterns and the efficacy of transit vehicles in accommodating downtown and long-distance commuting needs. Meanwhile, the study highlights the importance of striking an optimal balance between AMoD vehicle availability and transit service levels when designing the integrated urban mobility systems.

There are several limitations in this work. Firstly, it does not account for dynamic information within the commuter decision-making process. The discrete choice model applied assumes that commuter preferences for modes and routes are based on static information, an assumption that may not hold true in real-world scenarios. While real-time travel data for traditional transit systems pose a challenge, AMoD systems offer up-to-the-minute information on waiting times and estimated arrivals. Secondly, the model overlooks real-time traffic conditions, which can be significantly impacted by the introduction of AMoD vehicles, particularly in congested areas near subway stations. Lastly, the operational dynamics of AMoD systems, including vehicle rebalancing to redistribute vacant vehicles across different areas, are not considered. This aspect could notably enhance system efficiency. Addressing these limitations presents valuable avenues for future research.

\section*{Acknowledgment}
The authors would like to thank the MIT Energy Initiative for funding this research project. The authors would also like to thank Chicago Transit Authority (CTA) for offering data availability for this research.
% \newpage
\bibliographystyle{IEEEtranN}
\bibliography{IEEEabrv,reference}

% \newpage
% \input{appendix}

\end{document}